\documentclass[lettersize,journal]{IEEEtran}
\usepackage{amsmath,amsfonts}
\usepackage{algorithmic}
\usepackage{algorithm}
\usepackage{array}
\usepackage{textcomp}
\usepackage{stfloats}
\usepackage{url}
\usepackage{verbatim}
\usepackage{graphicx}
\usepackage{cite}
\usepackage{color}
\usepackage[utf8]{inputenc} 
\usepackage[T1]{fontenc}    
\usepackage{hyperref}      
\usepackage{url}            
\usepackage{booktabs}       
\usepackage{amsfonts}       
\usepackage{nicefrac}       
\usepackage{microtype}      
\usepackage{amsmath}
\usepackage{bbm}
\usepackage{amstext}
\usepackage{amssymb}
\usepackage{graphicx}
\usepackage{float}
\usepackage{subfigure}
\usepackage{multicol}
\usepackage{algorithm}
\usepackage{algorithmic}
\usepackage{wrapfig}
\usepackage{multirow}
\usepackage[table]{xcolor}
\usepackage{bigstrut}
\usepackage{graphicx}
\usepackage{amssymb}
\usepackage{diagbox}
\usepackage[normalem]{ulem}
\usepackage{ragged2e}
\makeatletter

\newcommand{\Rmnum}[1]{\expandafter\@slowromancap\romannumeral #1@}
\makeatother
\renewcommand\sout{\bgroup \color{green} \ULdepth=-.5ex \ULset}

\begin{document}

\title{
Investigating Deep Watermark Security: An Adversarial Transferability Perspective

}

\author{Biqing Qi\textsuperscript{1,2}, Junqi Gao\textsuperscript{3}, Yiang Luo\textsuperscript{1}, Jianxing Liu\textsuperscript{1}, ~\IEEEmembership{Senior Member,~IEEE,} Ligang Wu\textsuperscript{1}, ~\IEEEmembership{Fellow,~IEEE}, Bowen Zhou\textsuperscript{1,2},~\IEEEmembership{Fellow,~IEEE}
\thanks{

\textsuperscript{1}Department of Control Science and Engineering, Harbin Institute of Technology, Harbin, P. R. China; \textsuperscript{2}Department of Electronic Engineering, Tsinghua University, Bejing, P. R. China; \textsuperscript{3}School of Mathematics, Harbin Institute of Technology, Harbin, P. R. China. ( Emails: qibiqing7@gmail.com; gjunqi97@gmail.com; normanluo668@gmail.com; jx.liu@hit.edu.cn; ligangwu@hit.edu.cn; zhoubowen@tsinghua.edu.cn).

Corresponding authors: Bowen Zhou and Ligang Wu.
}

}

\markboth{Journal of \LaTeX\ Class Files,~Vol.~14, No.~8, August~2021}%
{Shell \MakeLowercase{\textit{et al.}}: A Sample Article Using IEEEtran.cls for IEEE Journals}


\maketitle

\begin{abstract}

The rise of generative neural networks has triggered an increased demand for intellectual property (IP) protection in generated content.
Deep watermarking techniques, recognized for their flexibility in IP protection, have garnered significant attention. However, the surge in adversarial transferable attacks poses unprecedented challenges to the security of deep watermarking techniques—an area currently lacking systematic investigation.
This study fills this gap by introducing two effective transferable attackers to assess the vulnerability of deep watermarks against erasure and tampering risks.
Specifically, we initially define the concept of local sample density, utilizing it to deduce theorems on the consistency of model outputs.
Upon discovering that perturbing samples towards high sample density regions (HSDR) of the target class enhances targeted adversarial transferability, we propose the Easy Sample Selection (ESS) mechanism and the  Easy Sample Matching Attack (ESMA) method.
Additionally, we propose the Bottleneck Enhanced Mixup (BEM) that integrates information bottleneck theory to reduce the generator's dependence on irrelevant noise.
Experiments show a significant enhancement in the success rate of targeted transfer attacks for both ESMA and BEM-ESMA methods. We further conduct a comprehensive evaluation using ESMA and BEM-ESMA as measurements, considering model architecture and watermark encoding length, and achieve some impressive findings.

\end{abstract}


\section{Introduction}

\IEEEPARstart{I}{n} recent years, deep learning-based generative models have advanced significantly, unlocking diverse applications like text-to-image and image-to-image to provide innovative user experiences \cite{56,57}. The implementation and deployment of these models have increased the urgency to address ethical security and copyright traceability in generated content. For example, companies using generative AI models aim to verify data ownership. They expect intellectual property (IP) protection measures to preserve the aesthetic appeal and user experience of the generated data. This introduces complexity that makes the design and development of secure IP protection technologies more challenging.

\begin{figure}[t]
  \centering
  \includegraphics[width=0.39\textwidth]{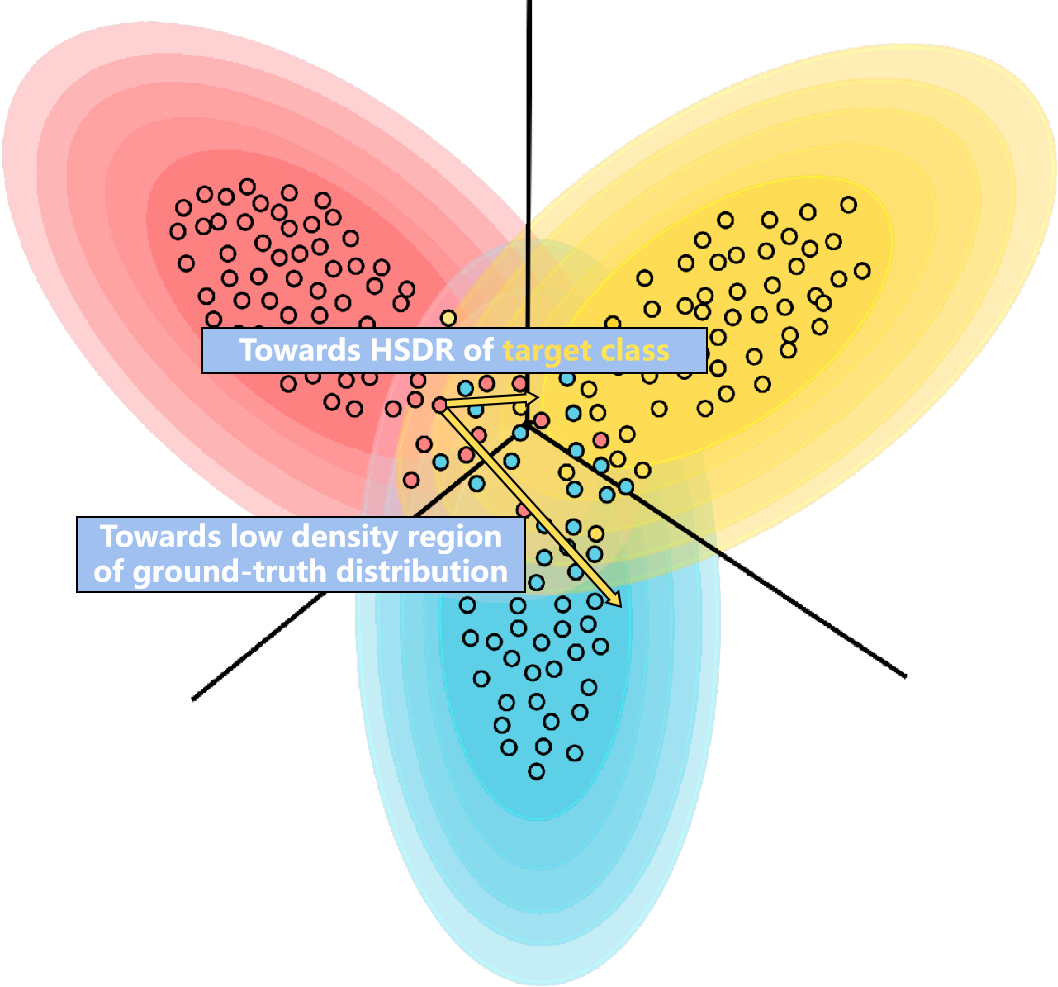} 
  \caption{A schematic example of our motivation, plotting the probability density (darker the color represents larger the density) and samples for three populations (orange, cyan, and green). The black line indicates the Bayesian discriminant boundary.}
  \label{fig.1}
\end{figure}

Deep watermarking is a common technology used for IP protection, encoding information within multimedia content for image labeling and copyright identification \cite{61, 75}. Deep watermarks, based on visual presentation, are classified as visible or invisible. Invisible watermarks use data redundancy for embedding, ensuring copyright authentication while maintaining visual aesthetics. 
Due to its superior visual experience while maintaining stable encoding and decoding, invisible watermarking is now the mainstream IP protection method for generative content.

Invisible watermarking relies on the characteristics of model redundancy \cite{66}, and a common strategy for watermark embedding involves the positive exploitation of adversarial sample generation technology. 
Recent studies \cite{5,6} have demonstrated the transferability of adversarial samples, naturally raising a critical concern: \textbf{are deep watermarks vulnerable to erasure or alteration via such transferable attacks}?
If so, attackers might erase or modify watermarks by training local proxy models using watermarks from various enterprises. However, the security of current deep watermarks has received little attention and lacks comprehensive evaluation.

Building upon the considerations and motivations mentioned above, this paper aims to explore potential risks and influencing factors of deep watermarking technology, specifically focusing on transferable attacks, viewed as a quantitative measurement.

Currently, transferable adversarial attacks can be classified into two categories: targeted attacks \cite{18,19} and non-targeted attacks \cite{11,12}, each associated with specific risks—erasure and tampering, respectively. Tampering, facilitated by targeted transfer attacks, necessitates the target model to misclassify samples from the ground-truth class into a specified class, thereby altering the original invisible watermark to a specified source.
In contrast, erasure, a result of non-targeted transferable attacks, involves the misclassification of samples from the ground-truth class into other classes, ultimately leading to the erasure of the original invisible watermark. 

For a comprehensive assessment, we consider both non-targeted and targeted attacks to evaluate the security of deep watermarking methods. In this evaluation, high-performance transferable attack techniques are crucial. To this end, we first introduce two high-performance transfer attack methods, applicable to both targeted and non-targeted attacks.
Subsequently, we analyze and evaluate the security of deep watermarking techniques based on this foundation.

Specifically, we reconsider the transferability of targeted attacks, focusing on the sample density. Previous research \cite{6} has shown that adversarial perturbations targeting low-density regions in the ground-truth distribution of the target class can improve adversarial transferability. 
However, in a targeted senario, we indicate that such a direction is not straightforward. 
As illustrated in Figure \ref{fig.1}, the low-density regions in the genuine distribution may not align with the locations of the target class.
Therefore, adversarial perturbations may not effectively execute targeted attacks. In contrast, perturbing the original samples to High Sample Density Regions (HSDR) of the target class provides a more straightforward choice. 
Through both theoretical and empirical verification, we confirm that diverse Deep neural networks consistently produce outputs in the HSDR of each class. This suggests that perturbing the HSDR of the target class not only offers a more direct path for targeted attacks but also enhances adversarial transferability.

However, the curse of dimensionality makes finding such samples impractical through density estimation in high-dimensional scenarios. To tackle this, using the concept of easy samples \cite{7}, we confirm that, for models using the early stopping mechanism, easy training samples with low loss are more likely to be in the HSDR. Based on this insight, we introduce the Easy Sample Matching Attack (ESMA) method.
It is a generative targeted attack strategy that utilizes Easy Sample Selection (ESS) to guide the generation of adversarial samples by identifying easy samples within the target class, thus facilitating targeted attacks.
To enhance its adaptability to source data and decrease the generator's dependence on data noise, we introduce Bottleneck-Enhanced Mixup (BEM) mechnism, leading to the BEM-ESMA method. 
Inspired by information bottleneck theory \cite{80}, BEM-ESMA integrates the mixup strategy \cite{76} to achieve both model robustness and data efficiency.

Our experiments on the ImageNet dataset show that ESMA and BEM-ESMA outperform current competitive benchmarks, including the state-of-the-art generative targeted attack TTP, achieving higher transfer success rates in targeted attacks.
Notably, BEM-ESMA significantly improves targeted adversarial transferability, showcasing exceptional effectiveness when applied across source and target models trained on distinct distribution data.

Building on the ESMA and BEM-ESMA methods, we carefully designe a series of progressive experimental setups to simulate scenarios involving transfer attacks, utilizing watermarked images from diverse sources. Our evaluation aims at comprehensively assessing the risks associated with deep watermarking through erasure and tampering transfer attack strategies. In particular, we extensively evaluate the security risk of various structured watermark models, such as HiDDeN \cite{66} based on the standard encoder-decoder structure, stability signature \cite{75} based on stable diffusion, and the flow-based watermark model FED \cite{78}, across different encoding bit lengths.

In summary, our contributions can be outlined as follows:

\begin{itemize}

\item We empirically and theoretically  demonstrate that deep learning models trained on the same distribution exhibit more consistent outputs within the HSDR for each class. Consequently, we confirm that adversarial perturbations towards the target class's HSDR can enhance targeted adversarial transferability.

\item To address the challenge of estimating samples within HSDR in high-dimensional scenarios, we empirically and theoretically confirmed that easy samples with low loss tend to be located within HSDRs. Based on this insight, we introduced the Easy Sample Matching Attack (ESMA), a generative transferable attack method.

\item We propose the Bottleneck Enhanced Mixup (BEM) mechanism, inspired by Information Bottleneck Theory, to enhance the generator's adaptability to data and improve transferability between source and target models trained on different data distributions. Experiments showed that ESMA equipped with the BEM mechanism (BEM-ESMA) achieved improved targeted adversarial transferability.

\item 
Our study, employing ESMA and BEM-ESMA across various model architectures and encoding lengths, unveils four critical vulnerabilities in deep watermarking:
1) Deep watermarking faces risks of erasure and tampering, and can even be manipulated to specific target watermarks. In specific scenarios, invisible deep watermarks are susceptible to non-targeted attacks and can be completely erased.
2) A trade-off exists between the transformation robustness and inherent security of deep watermarks.
3) Architectures with strong transformation robustness are more vulnerable to being targeted for tampering attempts.
4) A trade off exists between the encoding bits and adversarial robustness against erased attacks.

\end{itemize}

Compared to the preliminary conference version \cite{gao2023perturbation}, we have made significant improvements and expansions in this paper, mainly from 4 aspects:
1) We have rewritten the theoretical and experimental analysis sections in Sec \ref{sec 2}, enhancing the coherence and ease of following the argumentation process. 2) In Sec \ref{sec 2}, we introduced the BEM mechanism to further enhance ESMA, resulting in BEM-ESMA, which adapts training data distributions when the source model and target model are inconsistent.
3) In Sec \ref{sec 3}, leveraging our proposed ESMA and BEM-ESMA, we are the first to introduce a set of novel setups to explore the risk of erasure and tampering in image-based deep watermarking technologies.
4) In the experimental section, Sec \ref{sec 3}, we conducted comprehensive assessments of the security of deep watermarking technologies, evaluating them using ESMA and BEM-ESMA, and derived some impressive and meaningful findings.

\section{Related Works}

\subsection{Tranferable Adversarial Attacks}
Adversarial attacks can be divided into white-box attacks and black-box attacks. In the white-box setting, attackers have access to the model's structure and parameters, while in the black-box setting, they have no such information but only access to the input and output of the model \cite{qi2023improving}. This is typically the scenario encountered in real-world situations. Black-box attacks include query-based attacks \cite{8,9,10} and transfer-based attacks \cite{11,12,13}. Conducting too many queries is impractical in real-world applications. In contrast, transfer-based attacks are more feasible as they do not require queries. Transfer-based attacks often require the use of a white-box surrogate model to produce adversarial perturbations. In terms of the way adversarial perturbations are generated, there are two approaches: iterative instance-specific methods and generative methods. Iterative instance-specific methods utilize model gradients to iteratively add perturbations to specified samples (e.g., FGSM \cite{2}, C\&W \cite{4}, PGD \cite{14}). To enhance the transferability of adversarial samples, subsequent work combines these methods with various techniques, such as introducing momentum \cite{11,12} and considering input transformations \cite{12,13,15,56} during iterations, training auxiliary classifiers \cite{16,17}, or substituting different loss functions \cite{18,19,20}. However, instance-specific methods are primarily designed for non-targeted scenarios, often lacking effectiveness in targeted settings. Although relatively good results have been achieved for targeted attacks \cite{18,19,51}, instance-specific methods still require iteratively creating perturbations for each specified sample, while generators trained for generating adversarial perturbations can be generalized on more samples after training \cite{21,22,23,24}.

\subsection{Model Watermarking Methods}
The current deep watermarking methods have primarily two types of applications. One is model IP protection, which involves encoding watermarks within the model to prevent the risk of model theft by users. The other type is image copyright protection and tracing, where watermarks are directly encoded in images to ensure the ownership and origin of the disseminated images can be verified.

In terms of model IP protection, digital watermark embedding methods are mainly divided into white-box methods \cite{58, 59, 60, 61} and black-box methods \cite{62,63}. White-box methods directly embed watermark information in the model parameters or structure. Black-box methods do not directly modify the model itself and are suitable for third-party use or deployment of the model. They introduce specific patterns or markers in the training data and prove the ownership of the model's intellectual property by identifying the embedded watermark in the output results.

Compared to model IP protection, copyright protection and traceability of images have become more important due to the recent advancements in multimodal generative models such as GPT4 \cite{64} and DALLE \cite{65}, which allow users for creative freedom. Zhu et al. \cite{66} first introduced deep digital watermarking for performing invisible watermark steganography on images. This was the first end-to-end method based on DNN structures. By utilizing a noise-enhanced steganographic encoder, the normal decoding of the steganographic encoding can be guaranteed even in the presence of various image transformations such as cropping and JPEG compression. Ahmadi et al. \cite{67} proposed a deep end-to-end diffusion watermarking framework that embeds watermarks in the transformed domain while incorporating the idea of residuals. Building upon the works of Zhu et al. and Ahmadi et al., Liu et al. \cite{68} designed a redundant encoder-decoder model and proposed a novel two-stage separable deep learning framework to address the limitations of noise types and slow convergence in end-to-end training practices. Jia et al. \cite{69} presented the MBRS training method, which randomly selects either real JPEG, simulated JPEG, or no noise layer as the noise layer for different mini-batches, further enhancing the robustness of watermarks against JPEG compression distortions. To simultaneously ensure the robustness and invisibility of watermarks, Ma et al. \cite{71} first introduced the concept of Invertible Neural Networks (INN) into image watermarking techniques. Due to the natural compatibility between the inverse extraction process of INN and the decoder's norm, it avoids information loss caused by separately training the decoder in an end-to-end framework, thus avoiding the robust-invisible trade-off achieved by sacrificing imperceptibility to enhance robustness.

Due to the freedom and diversity of image generative models, the simple strategy of adding watermarks to existing instances encounters generalization issues when applied to image generation models, and is relatively inefficient. In order to make watermark computation in image generation models more efficient and secure, researchers have begun to explore directly adding watermarks to the models during the process of generating images. Wu et al. \cite{72} introduced watermark extraction supervision during the training of GANs, aiming to ensure that any image output by the model contains a specific watermark. Similarly, Fei et al. \cite{73} used a pre-trained decoder to supervise the training of GANs to ensure that the images generated by the model carry hidden watermarks. By detecting the watermarks in the output images, the ownership of the generation network can be verified. However, the increase in embedding capacity during the verification phase limits the image quality. To address this, Qiao et al. \cite{74} proposed a new GAN watermarking scheme, which generates watermarks in the generator through model training and trigger sets. During the process of verifying model ownership, the model owner can only trigger the watermark in the generator if the correct watermark label is obtained. In recent years, the Diffusion Model has demonstrated superior performance in image generation and has become a mainstream backbone model for AI painting and creation. Fernandez et al. \cite{75} first proposed incorporating watermarks into the generation process of Stable Diffusion (SD) and fine-tuning the SD decoder with a pre-trained watermark decoder, enabling it to decode images with watermarks based on latent space vectors.

\section{Methodology}
\label{sec 2}
In this section, we begin by investigating transfer attacks with a specific focus on sample density. Then, we provide a two-part analysis in \ref{subsec 21} to guide the development of our transfer attack strategy in \ref{subsec 22}. Finally, within \ref{subsec 23}, the Bottleneck Enhanced Mixup (BEM) mechanism is introduced to further reduce the dependency of the generator on irrelevant noise in the data, thereby enhancing the performance of ESMA, especially in the case that the source model and target model are trained on different distribution.

In the following sections, we use the following notations and definitions: $\mathcal Z = \mathcal X \times \mathcal Y\in \mathbb R^d\times \mathbb R$ is the sample space. The i.i.d. dataset $S=\left\{x_i,y_i\right\}_{i=1}^n$ consists of $n$ sample pairs $(x_i,y_i),1\le i \le n$. Given $y$, the conditional distribution of $x$ is $\mathcal D_{x\mid y}$. Feature mapping $f:\mathcal X\rightarrow \mathbb R^K$, where $K$ is the number of class, the class of feature mapping $f\in \mathcal F$. Specifically, the parametrized class $\mathcal F_{\boldsymbol w}:=\left\{f_{\boldsymbol w}:\mathcal X\rightarrow \mathbb R^K, \boldsymbol w\in \mathcal W \right\}$, where $\mathcal W$ is the parameter space. Define $\mathcal F_{\boldsymbol w}^{\mathcal S}:=\left\{f_{\boldsymbol w}^{\mathcal S}=\mathcal S\circ f_{\boldsymbol w}: f_{\boldsymbol w}\in \mathcal F_{\boldsymbol w}, \mathcal S\circ f_{\boldsymbol w}(x)=\mathcal S(f_{\boldsymbol w}(x))\right\}$, $\mathcal S$ denotes Softmax-transformation. The Softmax-Cross-Entropy loss is denoted as $\ell_{sce}(\cdot, \cdot): \mathbb R^K\times \mathbb R \rightarrow \mathbb R$. Let $S_j:=\left\{(x,y)\in S: y=j\right\}$, $\mathcal I_j:=\left\{i:(x_i,y_i)\in S^j \right\}$, $\mathcal C_j:=\left\{x\in \mathcal X: (x,y)\in \mathcal Z, y=j \right\}$. 

\newtheorem{definition}{Definition}
\begin{definition}[$(j,x_0,r)$-Local sample density] Given a class $j\in [K]$, $(x_0,y_0)\in S_j$, the $(j,x_0,r)$-Local sample density:
\begin{equation}
\rho_{(j,x_0,r)}=\frac{\sum_{i\in\mathcal I_j}\mathbbm{1}(x_i\in \mathcal B(x_0,r))}{\mathrm{vol}\mathcal B(x_0,r)}
\end{equation}
where $\mathcal B(x_0,r):=\left\{x\in \mathbb R^d:\|x-x_0\|\le r \right\}$, $\mathrm{vol}\mathcal B(x_0,r)$ denotes the volume of $\mathcal B(x_0,r)$.
\end{definition}
Now we use the notation $\mathcal I_{(j,x_0,r)}=\left\{i:(x_i,y_i)\in S_j,x_i\in \mathcal B(x_0,r),(x_0,y_0)\in S_j)\right\}$, $\mathcal C_{(j,x_0,r)}=\left\{x:x\in\mathcal C_j, x_i\in \mathcal B(x_0,r),(x_0,y_0)\in S_j)\right\}$. For simplicity, we denote $\ell_{sce}(f_{\boldsymbol w}(x),y)$ as $\ell(\boldsymbol w,x)$ and define the local empirical risk $R_{(j,x_0,r)}(\boldsymbol w)=\frac{1}{\left|\mathcal I_{(j,x_0,r)}\right|}\sum_{i\in\mathcal I_{(j,x_0,r)}}\ell(\boldsymbol w,x_i)$.
\subsection{Analytical Theory and Experiments About Adversarial Transferability}
\label{subsec 21}
\subsubsection{The Output Consistency in HSDR}
\begin{figure*}[t]
\centering  
\subfigure[]{
\includegraphics[width=0.232\textwidth]{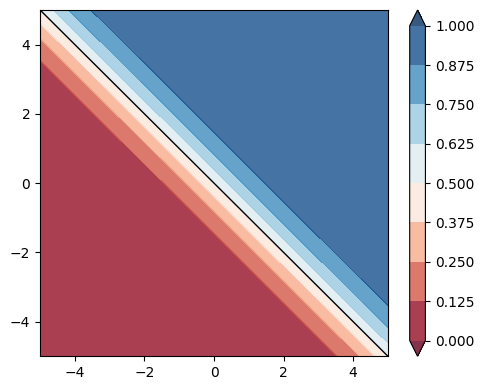}}
\subfigure[]{
\includegraphics[width=0.232\textwidth]{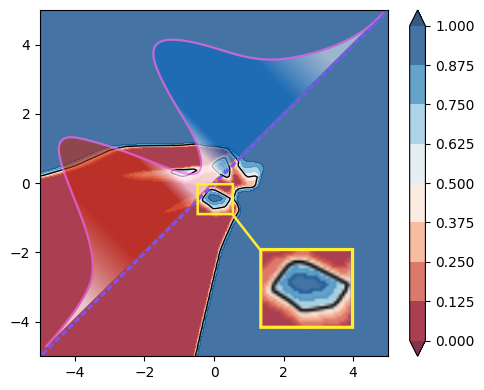}}
\subfigure[]{
\includegraphics[width=0.232\textwidth]{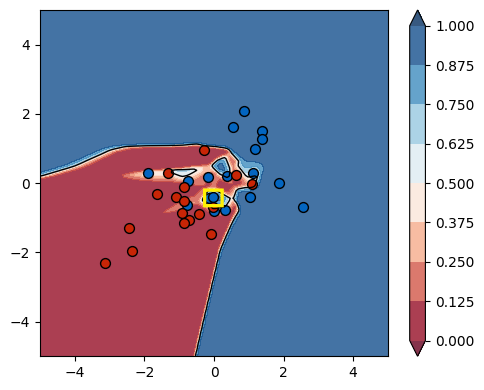}}
\subfigure[]{
\includegraphics[width=0.232\textwidth]{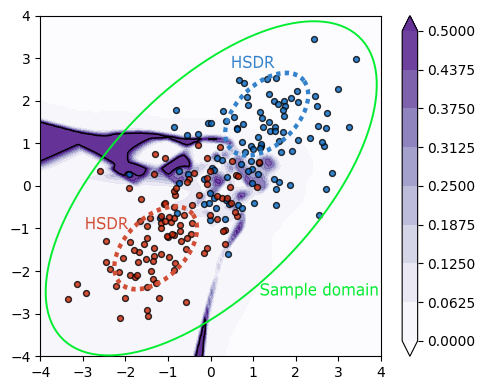}}
  \caption{(a): Bayesian discriminant region, darker the color indicate higher the confidence probability. (b): Classifier discriminant region, the probability density curves of the two population distributions are plotted, the white part represents the low-density region of ground-truth joint distribution, and we boxed out the small pits in the Bayesian misclassified region. (c): Classifier discriminant region with samples. We boxed an outlier. (d): Output differences between three different classifiers, darker purple indicates greater difference in output between different classifiers}
  \label{fig.2}

\end{figure*}

To enhance our understanding of adversarial transferability, we first analyze how local sample density affects model output consistency, then investigate the correlation between sample density and adversarial transferability. Specifically, if diverse models consistently generate similar outputs within a region, it suggests shared assessments in that area. Intuitively, perturbing original samples toward such a region should lead to perturbed samples demonstrating increased adversarial transferability across different models.
To verify this, we conducted a toy experiment to analyze how sample density impacts the consistency of model outputs. Specifically, we created a dataset consisting of $200$ samples by sampling from two $2$-D Gaussian distributions with equal probability. This dataset was then utilized to train a neural network for classification.
Next, we compare the discriminant region of the Bayes' criterion (minimizing error rate) for the known ground-truth prior distribution with that of the trained classifier.
Additionally, we train two neural networks with distinct structures and parameter quantities, and compare the output differences among the three networks (Figure \ref{fig.2}).

The discriminant region in Figure \ref{fig.2}(a) achieves the Bayesian error rate. Additionally, the intersection of the two population distributions at the center of the probability density curve (Figure \ref{fig.2}(b)) belongs to the low-density region of the ground-truth distribution and is also part of the misclassified region.
In this region, the Bayesian discriminant criterion, while minimizing expected error, might classify samples with relatively low probability density of a single population into another class, marking them as 'outliers'. The trained NN classifier distinguishes samples into their original categories, resulting in shifted decision boundaries compared to the Bayesian prior classifier, forming small pits as illustrated in Figure \ref{fig.2}(a).
Most of the samples surrounding the pit (Figure \ref{fig.2}(c)) belong to a class with higer population density compared to the class of samples in the pit. As a result, perturbing samples from this pit (i.e., outliers) toward the discriminant region of the other class becomes more achievable. If other trained neural networks can accurately classify such outliers, they will also generate similar pits. Adding perturbations towards these pits to samples from another class leads to their perturbation into such pits, facilitating the transfer of adversarial samples between different classifiers. This explanation aligns with the findings in \cite{6}.

However, as indicated in Figure \ref{fig.1}, perturbations directed towards the low-density regions of the ground-truth distribution are not sufficiently for targeted attacks. In contrast, perturbations towards the High Sample Density Region (HSDR) of the target class are more direct. Integrating this insight with Figure \ref{fig.2}(d), we note that different classifiers exhibit more consistent outputs in the HSDR.

Based on the results from the toy example and analysis, we draw the following conclusions: \textbf{Various deep learning models demonstrate more consistent outputs in the HSDR}. To offer a more rigorous theoretical foundation, we introduce Theorem \ref{theorem.1} to depict the correlation between the output consistency of different models and sample density, characterizing this observed phenomenon.

\newtheorem{theorem}{\bfseries Theorem}

\begin{theorem}[\bfseries Local output consistency]

\label{theorem.1}
For a target class $j\in\left[K\right]$, and two different parametrized class $\mathcal F_{{\boldsymbol w}_1}:=\left\{f_{{\boldsymbol w}_1}:{{\boldsymbol w}_1}\in{\mathcal W}_1 \right\}$, $\mathcal F_{{\boldsymbol w_2}}:=\left\{f_{{\boldsymbol w_2}}:{{\boldsymbol w_2}}\in{\mathcal W_2} \right\}$, assume that $\frac{1}{\left|\mathcal I_{(j,x_0,r)}\right|}\left|\sum_{i\in  \mathcal I_{(j,x_0,r)}}\left(f_{{\boldsymbol w}_1}^{\mathcal S_k}(x_i)-f_{{\boldsymbol w}}^{\mathcal S_k}(x_i)\right)\right|\le\gamma$, then for any sample $(x_0,y_0)\in S_j$, in the neighborhood $\mathcal B(x_0,r)$,  with probability at least $1-\delta$, the following holds:
\begin{equation}
\begin{aligned}
&\left\|\mathbb E_{x\sim \mathcal D_{x\mid y}}\left[f_{{\boldsymbol w}_1}^{\mathcal S}(x)-f_{{\boldsymbol w}_2}^{\mathcal S}(x)\mid x\in\mathcal C_{(j,x_0,r)}\right]\right\|_\infty
\\ &\le \mathcal O\left(\sqrt{\frac{Kd^{d/2}}{\rho_{(j,x_0,r)}2^dr^d}}\Lambda_{(j,x_0,r)}(d)+\sqrt{\frac{d^{d/2}\log(2K/\delta) }{\rho_{(j,x_0,r)}2^{d+1}r^d}}\right)+\gamma,
\end{aligned}    
\end{equation}
where $\Lambda_{(j,x_0,r)}(d)=\log ^{2}\left(r^d\sqrt{\rho_{(j,x_0,r)}}\right)$
\end{theorem}

\paragraph{Remark.}Theorem \ref{theorem.1} verifies that diverse models exhibit a more consistent output in the vicinity of samples in the High Sample Density Region (HSDR). Specifically, the difference between the outputs is bounded by $\tilde O\left(\sqrt{\frac{Kd^{d/2} }{\rho_{(j,x_0,r)}2^{d}r^d}}\right)$. When the number of classes and dimensions remains constant, higher sample density results in enhanced performance consistency, albeit with weaker consistency within smaller neighborhoods. It's important to note that this local consistency weakens as the dimension increases, but the relative output consistency between HSDR and Low Sample Density Region (LSDR) remains unchanged.

Hence, \textbf{perturbing original samples towards the target class enhances targeted adversarial transferability}. However, a practical challenge arises in high-dimensional spaces where sample points are often discrete due to the \emph{curse of dimensionality} \cite{26}. This complicates the determination of a suitable local neighborhood size for calculating sample density. Moreover, in the context of large datasets, computing local density becomes computationally expensive. Consequently, we must explore a more efficient strategy to identify the HSDR without the need for direct sample density calculations.

To address the aforementioned challenges, we propose an indirect approach centered on sample complexity. This choice is motivated by the significant attention given to hard samples across various tasks, as they play a crucial role in influencing training convergence and model generalization \cite{7,27,28,29}.
\cite{7} proposed that for a convergent model, the sample difficulty can be assessed using the loss gradient norm. Easy samples exhibit a relatively small loss gradient norm, whereas difficult samples have a comparatively large loss gradient norm. Moreover, those with an excessively large gradient norm may be considered outliers.
Inspired by this, we introduce the notions of hard and easy samples, framing the discussion within the context of model optimization, to establish a correlation between sample difficulty and sample density. Leveraging this connection, we streamline the process of identifying the HSDR by targeting easy samples with low loss. This circumvents the complexity of directly estimating HSDR through challenging density estimation. 

Next, we delve into model optimization, establishing a theoretical link between sample difficulty and sample density to facilitate analysis and insights. Prior to this, a few mild assumptions need consideration:

\newtheorem{assumption}{Assumption}
\begin{assumption}[Smoothness assumption]
\label{assumption1}
$\ell(\boldsymbol w,x)$ satisfies the following conditions of Lipschitz continuous gradient, for any $\boldsymbol w_1,\boldsymbol w_2\in\mathcal W$ and $ x_1,x_2\in\mathcal X$ separately:
\begin{equation}
\begin{aligned}
&\left\|\nabla_{\boldsymbol w}\ell(\boldsymbol w_1,x)-\nabla_{\boldsymbol w}\ell(\boldsymbol w_2,x)\right\|\le L_{1}\left\|\boldsymbol w_1-\boldsymbol w_2\right\|,\\
&\left\|\nabla_{x}\ell(\boldsymbol w,x_1)-\nabla_{x}\ell(\boldsymbol w,x_2)\right\|\le L_{2}\left\|x_1- x_2\right\|.
\end{aligned} 
\end{equation}
\end{assumption}
\begin{assumption}
\label{assumption2}
$\left\|\nabla_{\boldsymbol w}\ell(\boldsymbol w,x)\right\|\le G$ for all $\boldsymbol w\in \mathcal W$.
\end{assumption}
\begin{assumption}[Polyak-Łojasiewicz Condition]
\label{assumption3}
$R_{(j,x_0,r)}(\boldsymbol w)$ satisfies the PL-condition:
\begin{equation}
\frac{1}{2}\left\|\nabla_{\boldsymbol w}R_{(j,x_0,r)}(\boldsymbol w)\right\|^2\ge \mu\left(R_{(j,x_0,r)}(\boldsymbol w)-R_{(j,x_0,r)}^*\right),
\end{equation}
where $R_{(j,x_0,r)}(\boldsymbol w)=\frac{1}{\left|\mathcal I_{(j,x_0,r)}\right|}\sum_{i\in\mathcal I_{(j,x_0,r)}}\ell(\boldsymbol w,x_i)$ is the local empirical risk, $R_{(j,x_0,r)}^*=\underset{\boldsymbol w\in\mathcal W}{\inf}R_{(j,x_0,r)}(\boldsymbol w)$.
\end{assumption}
\begin{assumption}
\label{assumption4}
For $\forall i\in\mathcal I_{(j,x_0,r)}$, the following holds:
\begin{equation}
\left \langle \nabla_{\boldsymbol w}R_{(j,x_0,r)}(\boldsymbol w),\nabla_{\boldsymbol w}\ell(\boldsymbol w,x_i)\right \rangle \ge \beta\left \| \nabla_{\boldsymbol w}R_{(j,x_0,r)}(\boldsymbol w) \right \|^2.
\end{equation}
\end{assumption}
Assumptions \ref{assumption1} and \ref{assumption2} were posited in \cite{41}, \cite{42}, and \cite{43}. Notably, Assumption \ref{assumption2} can be derived from Assumption \ref{assumption1} in cases where the input space is bounded, a condition commonly met. Even non-convex functions can adhere to Assumption \ref{assumption3} \cite{42}; the inequality in Assumption \ref{assumption3} implies that all stationary points are global minima \cite{44}, a proof provided in recent works \cite{45,46} for over-parameterized DNNs. Assumption \ref{assumption4} holds reasonable validity when the local neighborhood radius $r$ is small.\par

\begin{algorithm}

\caption{Mini-batch SGD}          
\begin{algorithmic}[1]
\REQUIRE Initialized weights $\boldsymbol w^1$, total steps $T$, sample set $S$, batch size $M$ and step size $\eta_t$.
\FOR{$t=1\leftarrow T$}
\STATE Randomly sample $M$ different samples $B^t$ from $S$, where batch size $\left|B_t\right|=M$, corresponding indicator set denote as $\mathcal I_{B^t}$.
\STATE $\boldsymbol w^{t+1}=\boldsymbol w^{t}-\frac{1}{M}\sum_{i\in{\mathcal I_{B^t}}}\nabla_{\boldsymbol w}\ell(\boldsymbol w^t, x_i)$
\ENDFOR
\ENSURE $\boldsymbol w^{T+1}$
\end{algorithmic}

\label{algorithm 1} 
\end{algorithm}

In the context of local empirical risk optimization using Algorithm \ref{algorithm 1}, we present Theorem \ref{theorem.2} and Proposition \ref{proposition.1} to forge a theoretical link between sample difficulty and sample density.
\begin{theorem}[\bfseries Optimization relies on local density]
\label{theorem.2}
Given a parameterized class $\mathcal F_{\boldsymbol w}:=\left\{f_{\boldsymbol w}:\mathcal X\rightarrow \mathbb R^K, \boldsymbol w\in \mathcal W \right\}$, let $\boldsymbol w^t$ updated by Algorithm \ref{algorithm 1}, under Assumption \ref{assumption1}, \ref{assumption2}, \ref{assumption3} and \ref{assumption4}, set $\eta_t=\frac{1}{\beta\mu t}$ and $T\le\frac{L_1}{2\beta\mu}$, then with probability at least $1-\delta$, holds
\begin{equation}
\begin{aligned}
&R_{(j,x_0,r)}(\boldsymbol w^{t+1})-R_{(j,x_0,r)}^*\\
&\le\left ( 1-\frac{2}{t}\tau \left(\rho_{(j,x_0,r)}\right)  \right )  \left(R_{(j,x_0,r)}(\boldsymbol w^t)-R_{(j,x_0,r)}^*\right)+o\left ( \frac{1}{t^2} \right ),
\end{aligned}
\end{equation}
where $\tau \left(\rho_{(j,x_0,r)}\right)=\max\left \{ \left ( \frac{\rho_{(j,x_0,r)}\pi^{d/2}r^d}{\Gamma\left ( \frac{d}{2}+1 \right )M } -\sqrt{\frac{\ln\left ( T/\delta \right ) }{2M}} \right ),0 \right \}$, which is a non-descending function of $\rho_{(j,x_0,r)}$.
\end{theorem}

According to Theorem \ref{theorem.2}, the local empirical risk $R_{(j,x_0,r)}(\boldsymbol w)$ achieves a faster convergence rate in HSDR and a relatively slower convergence rate in LSDR. Consequently, early-stopping models exhibit lower local empirical risk in HSDR. When combined with the subsequent Proposition \ref{proposition.1}, we demonstrate that samples with smaller gradient norms and lower loss experience a reduced local empirical risk in their neighborhood.

\begin{figure*}[t]

\centering 
\subfigure{
\includegraphics[width=0.233\textwidth]{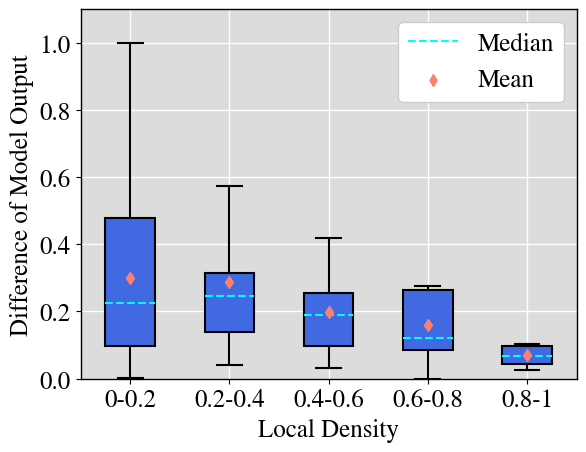}}
\subfigure{
\includegraphics[width=0.233\textwidth]{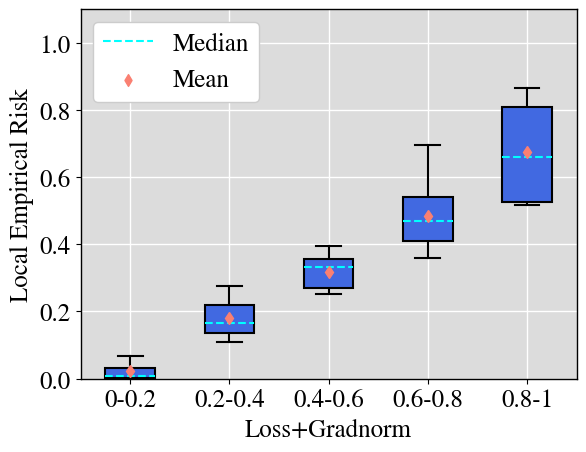}}
\subfigure{
\includegraphics[width=0.233\textwidth]{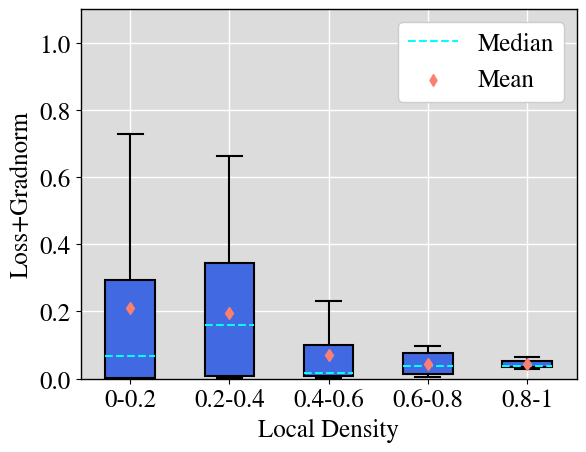}}
\subfigure{
\includegraphics[width=0.233\textwidth]{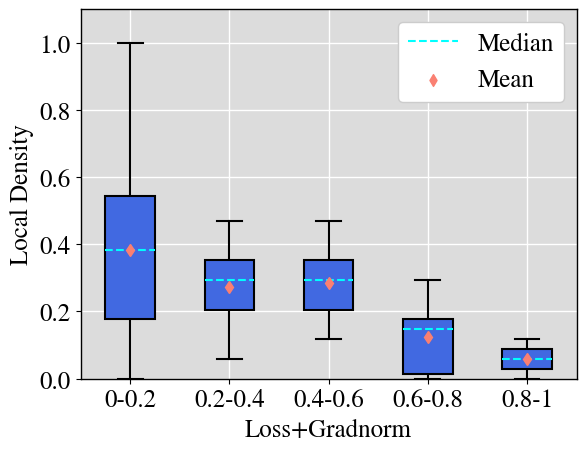}}

\caption{The first figure depicts the difference in output of three models under different local sample densities $\rho_{(y_i,x_i,r)}$ divided into different bins. The second figure shows the local empirical risk $R_{(y_i,x_i,r)}$ of samples under different sum of loss and gradient norms (\textbf{Loss+Gradnorm}). For Loss+Gradnorm, we first normalize both variables separately and then add them up to eliminate magnitude differences. The third figure represents the local empirical risk of local sample densities in different values. The fourth figure displays the local density under different Loss+Gradnorms. The neighborhood radius $r$ is taken as $0.4$.}
\label{Figure 3}

\end{figure*}

\newtheorem{proposition}{Proposition}
\begin{proposition}
\label{proposition.1}
Under Assumption \ref{assumption1}, given any $(x_i,y_i)\in S$, for any $x\in\mathcal X$ that satisfies $\left\|x_i-x\right\|\le r$, the following holds:
\begin{equation}
\frac{\Delta_{x_i}^{x}\ell_{\boldsymbol w}}{r}-\frac{3L_2r}{2} \le \left \| \nabla_x\ell\left ( \boldsymbol w,x_i \right ) \right \|\le\frac{\ell\left ( \boldsymbol w,x_i \right )}{r}+\frac{3L_2r}{2}, 
\end{equation} 
where $\Delta_{x_i}^{x}\ell_{\boldsymbol w} = \left | \ell\left ( \boldsymbol w,x_i \right )- \ell\left ( \boldsymbol w,x \right ) \right |$.
\end{proposition}

\paragraph{Remark.}
Proposition \ref{proposition.1} indicates that minimizing the loss results in a smaller norm of the loss gradient. However, this constraint weakens as the neighborhood radius $r$ decreases. Unlike the loss itself, the loss gradient norm provides an additional assurance of the proximity of local loss, particularly when $r$ is small. Hence, samples with smaller loss and a reduced loss gradient norm are more likely to reside in a neighborhood with a smaller local empirical risk.

For overfitting models, the local empirical risk in each neighborhood where samples are situated may be exceedingly small. However, for early-stopping models, this correlation between local empirical risk and local sample density remains intact.
Using the aforementioned theoretical analysis, for an early-stopping model, \textbf{samples with smaller loss and loss gradient norms are likely to be located in HSDR}.

To experimentally validate this, in the same example as the previous section, we train three models with early stopping, adjusting the learning rate with the number of steps. We then plotted Figure \ref{Figure 3}. The models exhibit more consistent outputs in HSDR. When the loss and gradient norms of a sample are small, the region where the sample is located has a smaller local empirical risk. Samples in HSDR have a smaller local empirical risk, aligning with our theoretical analysis.

Moreover, the rightmost graph in Figure \ref{Figure 3} indicates that samples with smaller loss and gradient norms are frequently situated in HSDR, confirming our discovery.

Hence, we can assess whether a sample is more likely to be located in HSDR or LSDR by checking if it possesses smaller loss and gradient norms simultaneously. This obviates the necessity to compute local sample densities for identifying samples in HSDR.

Combining the above analysis, we conclude that \textbf{adding perturbations towards easy samples with low loss of the target class can further enhance adversarial transferability}, based on this, we propose Easy Sample Screening (ESS, Algorithm \ref{algorithm 2}) to select anchor points for each target class to guide the addition of adversarial perturbations. To experimentally validate the effectiveness of ESS, we design the following comparative experiment. 
Specifically, we conducted transfer attack experiments on three baseline models using the CIFAR10 dataset, selecting three different models as victims. Our strategy involves using squared loss to match anchor points, minimizing $\left |f(x^{adv}i)- a{\text{target}_i} \right |^2$ (with $q=10$ and $\epsilon=16 \text{ pixels}$). For comparison, we used cross-entropy (CE) loss for traditional adversarial examples, and squared error without a screening mechanism (randomly selecting the target anchor in the target class) to mitigate the influence of different losses. The results are presented in Table \ref{tab1}, confirming that our strategy indeed enhances the transferability of target attacks, further validating our perspectives.
In addition, the detailed proof of the above theorems can be found in the appendix of our conference version\cite{gao2023perturbation}.

\begin{algorithm}[t]
\caption{The ESS Mechanism}          

\begin{algorithmic}[1]
\REQUIRE Early-stopping classifier $f_{\boldsymbol w}$, screening parameter $q$ and sample set $S$.
\FOR{$i=1\leftarrow n$}
\STATE{$\text{loss}_i=\ell_{sce}(f(x_i),y_i)$}
\STATE{$\text{gradnorm}_i=\|\nabla_{x}\ell_{sce}(f(x_i),y_i)\|$.}
\ENDFOR
\STATE{For each class $k\in [K]$, select the $q$-th smallest loss and the gradient norm among the samples in that class as thresholds $\text{thr}_k^{\text{loss}}$ and $\text{thr}_k^{\text{gradnorm}}$}
\FOR{$k=1\leftarrow K$}
\STATE{$A_k:=\left\{i\in \mathcal I_k:\text{loss}_i<\text{thr}_k^{\text{loss}},\text{gradnorm}_i<\text{thr}_k^{\text{gradnorm}}\right\}$},\par
$a_k = \frac{1}{\left|A_k\right|}\sum_{j\in A_k}f_{\boldsymbol w}(x_j)$:
\ENDFOR

\end{algorithmic}
\label{algorithm 2} 
\end{algorithm}

\definecolor{mybrown}{RGB}{166, 77, 88}

\begin{table*}[htbp]
  \centering
  \caption{Targeted transfer success rates. "left/middle/right" represent attacks using regular CE loss, square loss with randomly selected target anchors, and square loss with target anchor screening, respectively. In parentheses, the clean accuracy is indicated.}
  \renewcommand{\arraystretch}{1}
  \setlength{\tabcolsep}{1pt} 
    \begin{tabular}{c|ccccll}
    \toprule[1pt]
    \multirow{2}[1]{*}{Attack} & \multicolumn{2}{c}{\textbf{\textcolor{mybrown}{Src:Res34}}(95.44\%)} & \multicolumn{2}{c}{\textbf{\textcolor{mybrown}{Src:VGG16}}(94.27\%)} & \multicolumn{2}{c}{\textbf{\textcolor{mybrown}{Src:Dense121}}(95.47\%)} \\
      & $\rightarrow$VGG16 & $\rightarrow$Dense121 & $\rightarrow$Res34 & $\rightarrow$Dense121 & \multicolumn{1}{c}{$\rightarrow$Res34} & \multicolumn{1}{c}{$\rightarrow$VGG16} \\
    \midrule[0.5pt]
    MIM & 14.32\%/12.94\%/{\bfseries 14.44}\% & 19.81\%/19.00\%/{\bfseries 20.09}\% & 14.80\%/13.78\%/{\bfseries 15.13}\% & 13.41\%/12.11\%/{\bfseries 13.83}\% & 17.13\%/14.17\%/{\bfseries 17.17}\% & 11.23\%/10.27\%/{\bfseries 11.25}\% \\
    TIM & 13.53\%/12.50\%/{\bfseries 13.88}\% & 15.23\%/14.56\%/{\bfseries 15.71}\% & 12.49\%/11.78\% /{\bfseries 12.88}\% & 11.14\%/ 10.17\% /{\bfseries 11.40}\% & 15.74\%/14.61\%/{\bfseries 15.89}\% & 11.93\%/10.94\%/{\bfseries 12.09}\% \\
    DIM & 15.46\%/14.39\%/{\bfseries 15.97}\% & 18.10\%/ 17.61\%/{\bfseries 18.96}\% & 14.56\%/13.28\%/{\bfseries 15.12}\% & 12.97\%/12.11\%/{\bfseries 13.47}\% & 18.11\%/16.83\%/{\bfseries 18.33}\% & 13.18\%/11.72\%/{\bfseries 13.26}\% \\
    \bottomrule[1pt]
    \end{tabular}
  \label{tab1}

\end{table*}

\subsection{Construction of The Attack Strategy}
\label{subsec 22}
\subsubsection{Construction of ESMA}

\begin{figure}[h]

\centering 

\includegraphics[width=0.46\textwidth]{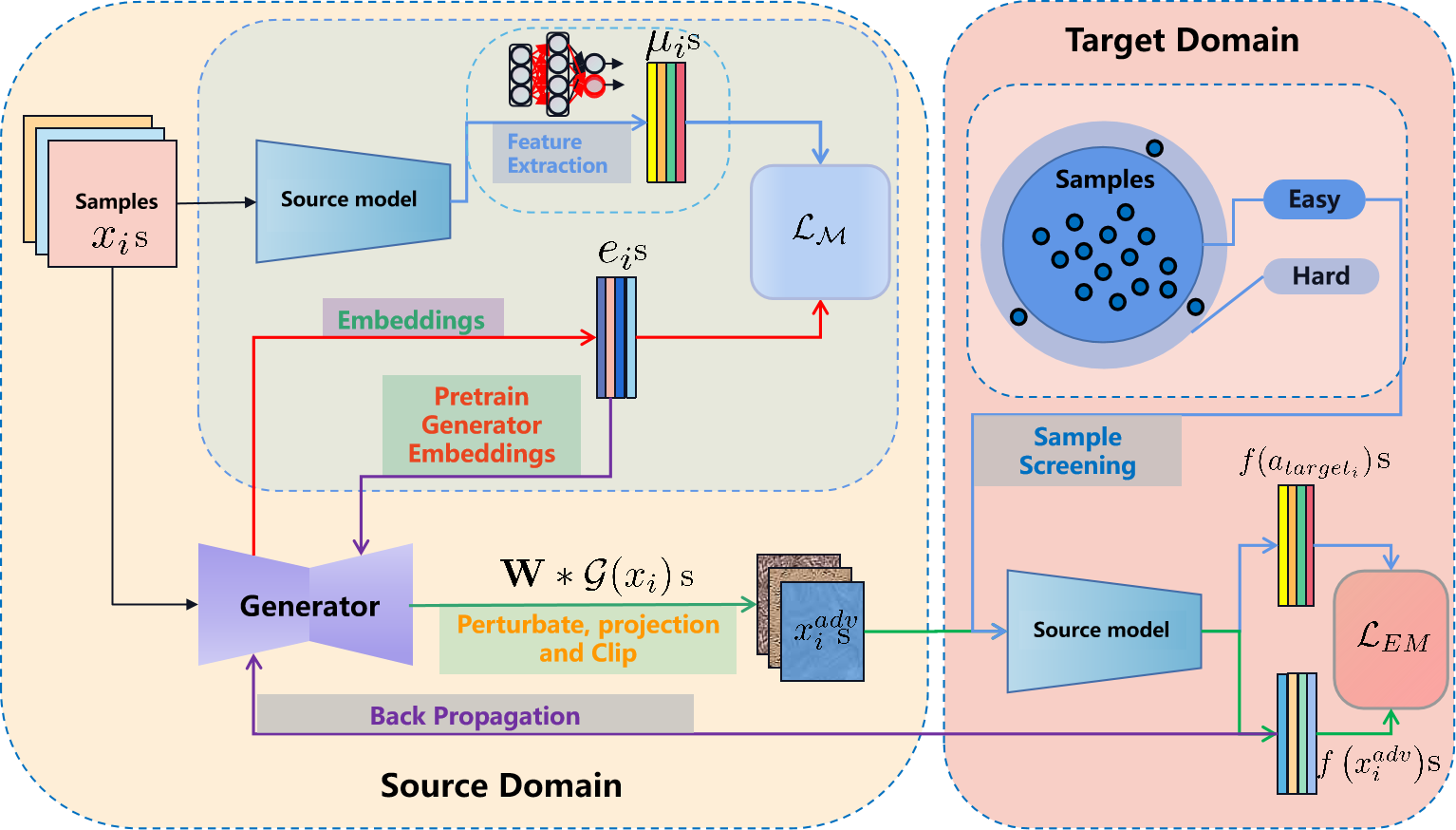}

\caption{Training strategy of ESMA.}
\label{Figure 4}

\end{figure}

Building upon the insights gained from the earlier findings, we formulate a generative targeted attack strategy termed Easy Sample Matching Attack (ESMA) leveraging the proposed ESS mechanism, illustrated in Figure \ref{Figure 4}. The training process involves two key steps. Initially, we pretrain the generator's embedding representations to facilitate multi-target attacks. Subsequently, in the second step, we employ our proposed ESS mechanism to identify easy samples for each target class. These selected samples guide the generator in creating adversarial examples tailored for targeted attacks.

\paragraph{Pre-trained Embeddings Guided by Latent Features}
To pretrain a well-structured set of class embeddings for the generator, ensuring it can effectively execute targeted attacks on specified classes without confusion, we leverage the insight that the latent space of deep learning models often contains ample class information \cite{30}.
We consider the output features of a surrogate model as a set of pre-existing embeddings that guide the construction of class embeddings for the generator. To ensure that the structure of the generator's embeddings resembles these pre-existing embeddings, we adopt the concept of manifold learning, employing a strategy akin to the SNE algorithm \cite{25}. Let $\mu_j =\frac{1}{\left|\mathcal I_j\right|}\sum_{i\in \mathcal I_j}l_i$, where $l_i=f(x_i)$, and then we formulate the following manifold matching loss. It involves pulling the generator embeddings of different classes towards the manifold formed by the output features, resulting in embeddings with an improved structure.
\par
\begin{figure*}[t]

\centering 
\subfigure{
\includegraphics[width=0.45\textwidth]{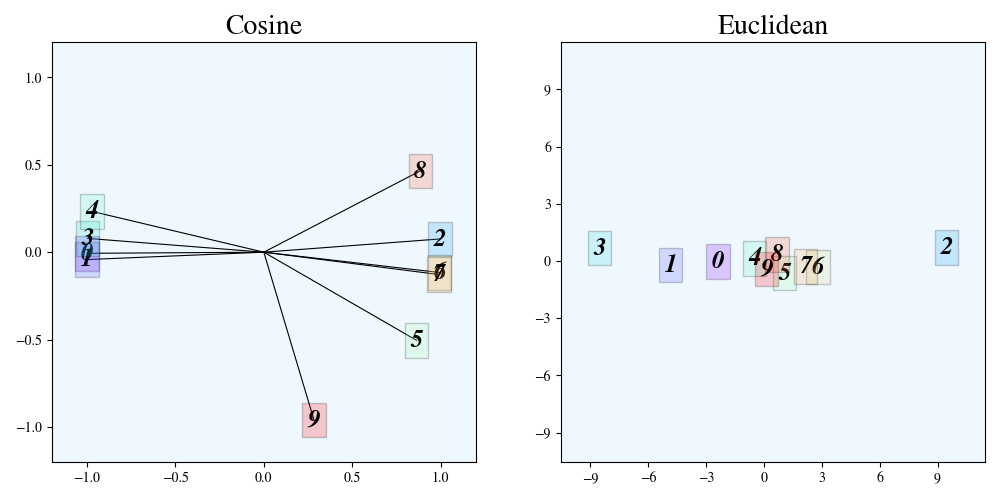}}
\subfigure{
\includegraphics[width=0.45\textwidth]{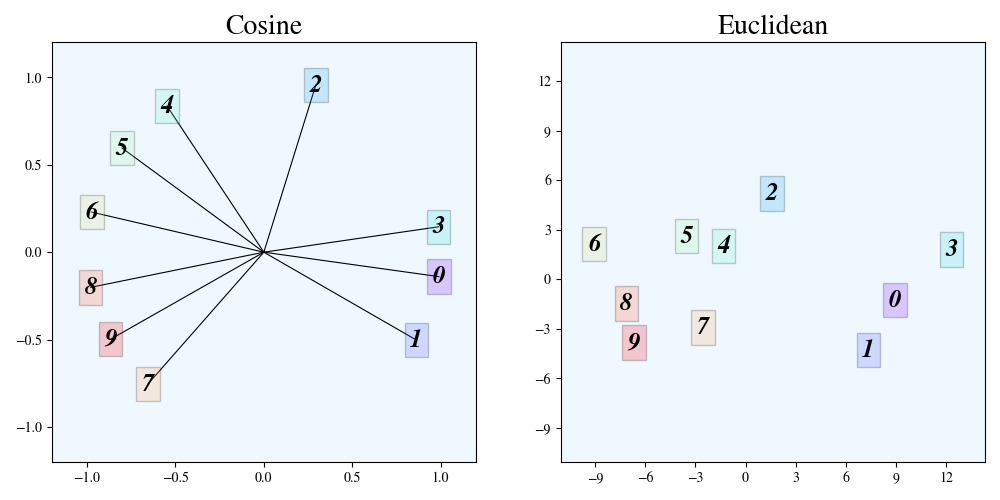}}
\caption{The two figures on the left showcase the outcomes of direct training for $10$ epochs, whereas the two figures on the right depict the results obtained by optimizing $\mathcal L_{\mathcal M}$. The numbers $0-9$ correspond to different class labels in sequential order. Each figure displays two images, representing distances after PCA projection onto a $2$-dimensional plane (normalized and unnormalized). The visualizations reflect cosine similarity and Euclidean distance, respectively.}
\label{Figure 5}

\end{figure*}

Next, we denote the generator embedding of class $j$ as $e_j$. The four matrices $M^{S_{euc}}$, $M^{E_{euc}}$, $M^{S_{cos}}$, $M^{E_{cos}}$ satisfy $M^{S_{euc}}_{i,j}=\|\mu_i-\mu_j\|$, $M^{E_{euc}}_{i,j}=\|e_i-e_j\|$, $M^{S_{cos}}_{i,j}=\frac{\mu_i\cdot\mu_j}{\|\mu_i\|\|\mu_j\|}$, $M^{E_{cos}}_{i,j}=\frac{e_i\cdot e_j}{\|e_i\|\|e_j\|}$, respectively. Let 
\vspace{-3pt}
\begin{equation}
\begin{array}{c}
\overline{M}_{i, j}^{S_{\text {euc }}}=\frac{\exp \left(M_{i, j}^{S_{\text {euc }}}\right)}{\sum_{k=1}^{K} \exp \left(M_{i, k}^{S_{\text {euc }}}\right)}, \overline{M}_{i, j}^{E_{\text {euc }}}=\frac{\exp \left(M_{i, j}^{E_{\text {euc }}}\right)}{\sum_{k=1}^{K} \exp \left(M_{i, k}^{E_{\text {euc }}}\right)}, \\
\overline{M}_{i, j}^{S_{\text {cos }}}=\frac{\exp \left(M_{i, j}^{S_{\text {cos }}}\right)}{\sum_{k=1}^{K} \exp \left(M_{i, k}^{S_{\text {cos }}}\right)}, \overline{M}_{i, j}^{E_{\text {cos }}}=\frac{\exp \left(M_{i, j}^{E_{\text {cos }}}\right)}{\sum_{k=1}^{K} \exp \left(M_{i, k}^{E_{\text {cos }}}\right)},
\end{array}
\end{equation}
then our manifold matching loss is as follows:
\begin{equation}
\begin{aligned}
\mathcal{L}_{\mathcal{M}}&=\sum_{i, j} \overline{M}_{i, j}^{S_{\text {euc }}} \log \frac{\overline{M}_{i, j}^{S_{\text {euc }}}}{\overline{M}_{i, j}^{E_{\mathrm{euc}}}}+\sum_{i, j} \overline{M}_{i, j}^{E_{\text {euc }}} \log \frac{\overline{M}_{i, j}^{E_{\mathrm{euc}}}}{\overline{M}_{i, j}^{S_{\mathrm{euc}}}} \\
&+\lambda_1\left[\sum_{i, j} \overline{M}_{i, j}^{S_{\mathrm{cos}}} \log \frac{\overline{M}_{i, j}^{S_{\mathrm{cos}}}}{\overline{M}_{i, j}^{E_{\mathrm{cos}}}}+\sum_{i, j} \overline{M}_{i, j}^{E_{\mathrm{cos}}} \log \frac{\overline{M}_{i, j}}{\overline{M}_{i, j}}\right]\\
&+\lambda_2\sum_{i=1}^{K}\left\|e^{i}\right\|.
\end{aligned}
\end{equation}
The last regularization term aims to prevent the collapse of losses, with $\lambda_1$ and $\lambda_2$ serving as hyperparameters. The pre-trained embeddings, utilizing our strategy, exhibit a larger Euclidean distance and a smaller cosine similarity. This significantly mitigates the clustering phenomenon observed previously, as illustrated in Figure \ref{Figure 5}.

\paragraph{Training of Multi-target Adversarial Perturbation Generators}
After pretraining the embeddings, we freeze the parameters of the embedding layer. Combined with our proposed ESS, we screen $q$ easy samples in each class to create target anchor sets $A_i$. Subsequently, we align these anchor sets with the output of our generator in feature space. The generator employed is a Unet with Resblocks. We introduce the following easy sample feature matching loss to train the multi-class adversarial generator.
\begin{equation}
\mathcal{L}_{EM}=\sum_{j=1} ^{K} \frac{1}{\sum_{i \in [n]} \mathbbm{1}\left(i \notin \mathcal I_{j}\right)} \sum_{i \notin \mathcal I_{j}}d\left ( a_{j},f\left(\operatorname{clip}_{\mathbf W,\epsilon}^{ \mathcal{G}_{\theta}\left(x_{i},y_{j}\right)}\right) \right ),
\end{equation}
where $\operatorname{clip}_{\mathbf W,\epsilon}^{ \mathcal{G}_{\theta}\left(x_{i},y_{j}\right)} = \operatorname{clip}_{\epsilon}\left(\mathbf W \ast \mathcal G\left(x_i,y_{j}\right)\right)$ and $\operatorname{clip}_{\epsilon}(x)= \operatorname{clip}\left(\min \left(x+\epsilon, \max \left(x, x-\epsilon\right)\right)\right)$, $\mathbf W$ is a differentiable Gaussian kernel with size $3\ast 3$, such smoothing operations have been demonstrated to further improve transferability\cite{21}. The measure of the distance between the two features $d(\cdot, \cdot)$ is Smooth L1 loss, which has a unique optimal solution and is not sensitive to exceptional values \cite{48}.

\subsubsection{Bottleneck-Enhanced Mixup}
\label{subsec 23}

To expand the applicability of ESMA beyond situations where the original and target domains perfectly align, we need to address the generator's dependence on data noise. This is critical because noise often manifests as non-invariant features, which are unstable and not tightly linked to the class information we want to capture. By focusing on extracting more general, invariant features, we inherently reduce the influence of non-invariant noise, leading to a more robust and transferable ESMA.

The Information Bottleneck (IB) theory\cite{80,81} serves as a natural framework for addressing this need. Inspired by the IB theory, we designate the random variables representing the original image, the adversarial sample, and the adversarial hidden features output by the model $f_{\boldsymbol w}$ as $X$, $X_{adv}$, and $Z_{adv}^{\boldsymbol w}$, respectively.
Considering the information flow in adversarial attacks: $X \rightarrow X_{adv} \rightarrow Z_{adv}^{\boldsymbol w}$, our objective is to optimize the adversarial generator loss $\mathcal L_{EM}$ while adhering to the constraint of the conditional mutual information between the original data $X$ and the adversarial sample $I(X,X_{adv}|Y_{tar})$, where $Y_{tar}$ represents the random variable of the target label. This method allows us to minimize the reliance of the adversarial sample on noise in the original data and enhance its dependence on invariant features. The conditional mutual information can be expressed in two different ways:
\begin{equation}
\begin{aligned}
&I(X;X_{adv}, Z_{adv}^{\boldsymbol w} | Y_{tar})\\
&=I(X; Z_{adv}^{\boldsymbol w} | Y_{tar}) +I(X; Z_{adv}^{\boldsymbol w} | Y_{tar}, X_{adv})\\
&=I(X; X_{adv} | Y_{tar}) +I(X; X_{adv} | Y_{tar}, Z_{adv}^{\boldsymbol w}).
\end{aligned}
\end{equation}
Due to Markovity, $X$ and $Z_{adv}^{\boldsymbol w}$ are independent. Then, based on the non-negativity of mutual information, we have:
\begin{equation}
I(X, X_{adv} | Y_{tar}) \ge I(X, Z_{adv}^{\boldsymbol w} | Y_{tar}), 
\end{equation}
thereby due to the fact that 
\begin{align}
&I(X, Z_{adv}^{\boldsymbol w} | y_{tar})\\
&=\int p({z}_{adv}^{\boldsymbol w}, \nonumber x,y_{tar})\log\frac{p({z}_{adv}^{\boldsymbol w}| x,y_{tar})}{p({z}_{adv}^{\boldsymbol w}|y_{tar})}d{z}_{adv}^{\boldsymbol w}dy_{tar}d x\\
&=\mathbb E_{X,y_{tar}}\left[\mathcal D_{\mathbf{KL}}(p({z}_{adv}^{\boldsymbol w}| x,y_{tar})\Vert p({z}_{adv}^{\boldsymbol w}|y_{tar})) \right],
\end{align}
Thus, we can minimize $\mathcal D_{\mathbf{KL}}(p({z}{adv}^{\boldsymbol w}| x,y{tar})\Vert p({z}{adv}^{\boldsymbol w}|Y{tar}))$ for any $\boldsymbol x\in\mathcal X$ to optimize the lower bound of $I(X, X_{adv} | y_{tar})$.
Therefore, we seek $p({z}{adv}^{\boldsymbol w}| x,y{tar})$ to be as close as possible to $p\left( z_{adv}^{\boldsymbol w}|y_{tar}\right)$. Setting $p\left( z_{adv}^{\boldsymbol w}|y_{tar}\right)$ as a one-point distribution located at $a_{y_{tar}}$, we propose the \textbf{Bottleneck Enhanced Mixup} (BEM), introducing these concepts into the training framework of ESMA to further enhance the generator’s adaptability to source data.
\begin{figure}

\centering  

\includegraphics[width=0.5\textwidth]{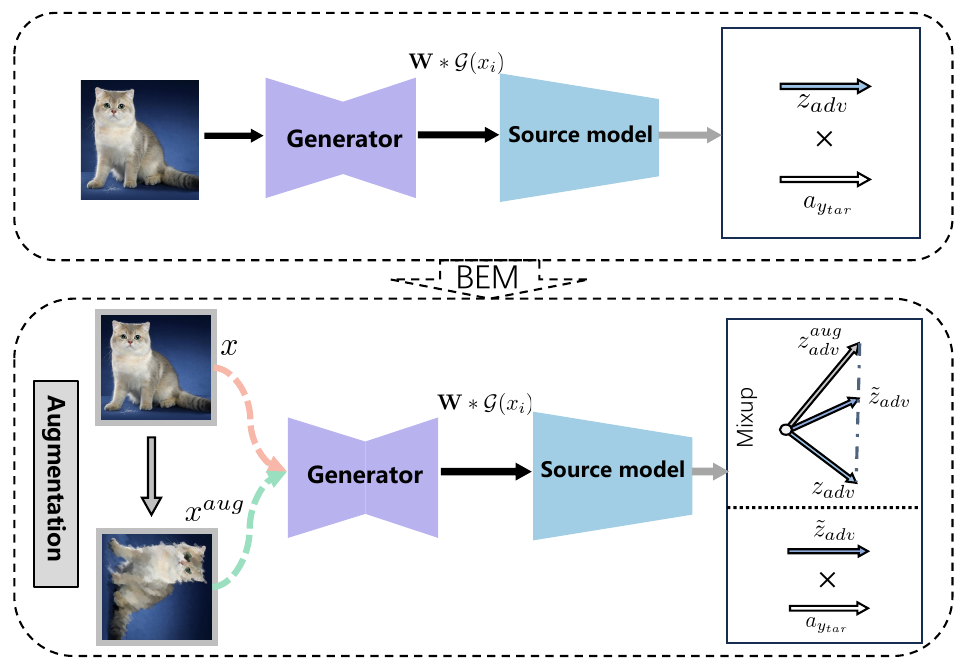}

\caption{Schematic diagram of BEM. Encouraging the vectors in the feature space, resulting from Mixup of randomly augmented images after being attacked and the original images after being attacked, to closely align with the anchor points of the target class ensures the stability of the generator's attacks amidst diverse data. This further diminishes the generator's reliance on the source data.}
\label{illustration}

\end{figure}
The mechanism of BEM is shown in Figure \ref{illustration}. By modeling ${\tilde z}_{adv}^{\boldsymbol w}=\zeta{z}_{adv}^{\boldsymbol w}+(1-\zeta){z}_{adv}^{aug,\boldsymbol w}$, where $\zeta\sim\text{Beta}(\nu,\nu)$, $z^{aug,\boldsymbol w}_{adv}=f_{\boldsymbol w}\left (\mathbf W\ast\mathcal G(x^{aug}, y_{tar})\right)$, $x^{aug}$ is the sample after data enhancement, and the Mixup distribution density $p({\tilde z}_{adv}^{\boldsymbol w}| x,y_{tar})= \zeta p({ z}_{adv}^{\boldsymbol w}| x,y_{tar})+\left(1 -\zeta\right)p({ z}_{adv}^{aug,\boldsymbol w}| x,y_{tar})$. It can be seen from the non-negativity of KL divergence that $\mathbb E_{X,y_{tar}}\left[\mathcal D_{\mathbf{KL}}(p({z}_{adv}^{\boldsymbol w} | x,y_{tar})\Vert p({\tilde z}_{adv}^{\boldsymbol w}|y_{tar})) \right]=0$ iff $p({ z}_{adv}^ {\boldsymbol w}| x,y_{tar}) = p({ z}_{adv}^{aug,\boldsymbol w}| x,y_{tar}) = p\left( z_{adv}^{\boldsymbol w}|y_{adv}\right)$ is established with probability $1$, this weakens the generator's dependence on irrelevant noise and further ensures that the generator can adapt to a wider range of source domain data. Note that the above properties only need $d(\cdot,\cdot)\ge 0$ and $d(a, b)=0$ iff $a=b$ to be guaranteed, so we can replace $f_{\boldsymbol w}\left(\operatorname{clip}_{\mathbf W,\epsilon}^{ \mathcal{G}_{\theta}\left(x_{adv}^{aug},y_{tar}\right)}\right) $ in $\mathcal L_{EM}$ with $\tilde z^{aug,\boldsymbol w}_{adv}$ to incorporate BEM into the ESMA training process:
\begin{equation}
\mathcal{L}_{B-EM}=\sum_{j=1} ^{K} \frac{1}{\sum_{i \in [n]} \mathbbm{1}\left(i \notin \mathcal I_{j}\right)} \sum_{i \notin \mathcal I_{j}}d\left ( a_{j},\tilde z^{aug,\boldsymbol w}_{adv} \right ).
\end{equation}

Then, our final training strategy of BEM-ESMA can be represented by algorithm \ref{algorithm 3}.

\begin{algorithm}
\caption{Training Strategy of BEM-ESMA}          
\begin{algorithmic}[1]
\REQUIRE Generator $\mathcal G_\theta$ with pre-trained embeddings, anchors $a_k, k\in[K]$ and Total epochs $N$.

\FOR{$t=1\leftarrow N$}
\FOR{$i=1\leftarrow n$}
\STATE{$\text{target}_i\sim \text{Uniform}(\{1,\dots,K\})$  }.
\IF {$y_i \ne \text{target}_i$}
\STATE{Random choice an anchor $a_j$ from $A_j$},
\STATE{Randomly sample a $\zeta_i\sim \text{Beta}\left(\nu, \nu\right)$,}
\STATE{Perform BEM operation ${\tilde z}_{adv}^{\boldsymbol w}=\zeta{z}_{adv}^{\boldsymbol w}+(1-\zeta){z}_{adv}^ {aug,\boldsymbol w}$.}
\ELSE
\STATE{Remove $x_i$ from batch $t$.}
\ENDIF
\ENDFOR
\STATE{Gradient descent step on batch wise $\mathcal L_{B-EM}$}.
\ENDFOR

\end{algorithmic}
\label{algorithm 3} 

\end{algorithm}

\section{Experiments}

\label{sec 3}
\subsection{Targeted Attack Transferability Evaluation}
\subsubsection{Implementation Details}

\paragraph{Dataset Setup}

The dataset used in this study is derived from the ILSVRC2012 dataset \cite{31}. Specifically, we choose ten classes from it as the training set, labeled as $24$, $99$, $198$, $245$, $344$, $471$, $661$, $701$, $802$, $919$, aligning with the classes used for TTP training in \cite{21}. Generators are trained on images of these ten classes within the training set, with $1300$ images per class. The images from the validation set, consisting of $50$ images per class, serve as the test dataset for the targeted attack.

\begin{table*}[htbp]
\scriptsize
  \centering
  \caption{Targeted transfer success rates. "Src" indicates the source model, the best results are bolded and the second are underlined.}
  \setlength{\tabcolsep}{7.5pt}
  \renewcommand{\arraystretch}{0.55} 
    \begin{tabular}{c|c|ccccccccc}
    \toprule[1pt]
    \textbf{Src}& Attack
      & →VGG19bn & →Dense121 & →Res152 & →Inc-v3 & →Inc-v4 & →IncRes-v2 & →ViT & AVG\\
    \cmidrule{1-10}
   \multirow{13}{*}{\centering{\textbf{\textcolor{mybrown}{Res50}}}}& MIM & 1.33\% & 3.44\% & 3.67\% & 0.51\% & 0.31\% & 0.22\% & 0.16\% & 1.38\%\\
   & TIM & 13.07\% & 26.16\% & 23.82\% & 5.51\% & 3.84\% & 2.96\% &  0.60\% & 10.85\% \\
   & DIM & 13.36\% & 24.96\% & 23.44\% & 5.11\% & 3.80\% & 2.29\%  & 0.47\% & 10.49\% \\
   & SI-NIM & 1.07\% & 1.91\% & 2.00\% & 0.40\% & 0.44\% & 0.29\%  & 0.07\% & 0.88\% \\
   & Po-TI-Trip & 17.93\% & 33.62\% & 32.20\% & 7.73\% & 5.69\% & 3.38\% & 1.07\% & 14.52\% \\
&	$\text{SI-FGS}^2\text{M}$ & 20.75\% & 32.68\% & 30.55\% & 6.30\% & 4.42\% & 2.92\% & 0.91\% & 14.08\% \\
&   $\text{S}^2$I-SI-TI-DIM & 26.49\% & 36.36\% & 36.49\% & 14.87\% & 12.67\% & 10.02\% & 1.18\% & 19.72\% \\
&	Logit & 51.60\% & 77.22\% & 76.98\% & 17.62\% & 12.67\% &8.40\% & 3.09\% & 35.37\%\\
&	DTMI-Logit-SU & 53.36\% & 78.91\% & 79.00\% & 18.56\% & 13.36\% &8.71\% & 3.32\% & 36.46\%\\
&	RAP-LS & 60.04\% & 83.40\% & 80.38\% & 22.16\% & 17.24\% & 10.82\% & 3.80\% & 39.69\% \\
&	HGN & 63.92\% & 70.64\% & 68.43\% & 22.41\% & 13.76\% & 7.31\% & 12.12\% & 36.94\% \\
 &   TTP & 67.11\% & 73.67\% & 72.64\% & 33.49\% & 25.27\% & 11.27\% & 18.07\% & 43.07\% \\
    \rowcolor{gray!20} &ESMA (ours) & \underline{81.29}\% & \underline{83.89}\% & \underline{81.78}\% & \underline{37.53}\% & \underline{39.29}\% & \underline{17.29}\% & \underline{21.25\%} & \underline{51.76\%} \\
    \rowcolor{gray!20} &BEM-ESMA (ours) & \textbf{83.52}\% & \textbf{84.11}\% & \textbf{84.25}\% & \textbf{38.12}\% & \textbf{40.13}\% & \textbf{17.81}\% & \textbf{22.30\%} & \textbf{52.89\%} \\
    \midrule[0.8pt]
   \textbf{Src} & {Attack} 
      & →VGG19bn & →Res50 & →Res152 & →Inc-v3 & →Inc-v4 & →IncRes-v2 & →ViT & AVG\\
    \cmidrule{1-10}
  \multirow{13}{*}{\centering{\textbf{\textcolor{mybrown}{Dense121}}}} & MIM & 1.98\% & 2.33\% & 1.51\% & 0.56\% & 0.40\% & 0.27\% & 0.27\% & 1.05\% \\
   & TIM & 8.98\% & 12.82\% & 8.87\% & 3.44\% & 3.31\% & 1.91\% & 0.71\% & 5.72\% \\
   & DIM & 10.20\% & 13.80\% & 9.18\% & 3.98\% & 3.27\% & 2.29\% & 0.60\% & 6.19\% \\
   & SI-NIM & 8.29\% & 11.42\% & 6.96\% & 1.78\% & 1.71\% & 1.13\% & 0.29\% & 4.51 \\
   & Po-TI-Trip & 11.29\% & 16.80\% & 11.78\% & 5.49\% & 4.84\% & 2.96\% & 0.96\% & 7.73\% \\
&	$\text{SI-FGS}^2\text{M}$ & 12.55\% & 16.67\% & 11.70\% & 5.16\% & 4.89\% & 3.58\% & 0.85\% & 7.91\% \\
&   $\text{S}^2$I-SI-TI-DIM & 25.60\% & 30.33\% & 23.98\% & 13.11\% & 12.04\% & 7.93\% & 1.38\% & 16.33\% \\
&	Logit & 31.98\% & 43.49\% & 31.71\% & 12.91\% & 11.51\% & 7.16\% & 3.27\% & 20.29\%\\
&	DTMI-Logit-SU & 33.38\% & 44.93\% & 34.16\% & 13.44\% & 12.36\% & 7.75\% & 3.62\% & 21.38\%\\
&	RAP-LS & 38.56\% & 50.67\% & 38.24\% & 15.78\% & 13.96\% & 9.64\% & 3.82\% & 24.38\% \\
&	HGN & 47.81\% & 56.96\% & 44.21\% & 22.75\% & 19.56\% & 9.42\% & 10.62\% & 30.19\% \\
 &   TTP & 52.00\% & 58.02\% & 49.24\% & 29.69\% & 23.00\% & 13.24\% & 17.71\% & 34.70\% \\
   \rowcolor{gray!20} & ESMA (ours) & \underline{56.83}\% & \textbf{63.71}\%  & \underline{54.70}\%  & \textbf{33.23}\%  & \underline{29.28}\%  & \textbf{15.97}\% & \textbf{18.50}\% & \underline{38.89}\% \\
   \rowcolor{gray!20} &BEM-ESMA (ours) & \textbf{58.36}\% & \underline{63.25}\% & \textbf{55.12}\% & \underline{32.65}\% & \textbf{30.67}\% & \underline{15.72}\% & \underline{17.61\%} & \textbf{39.05\%} \\
    \midrule[0.8pt]
    \textbf{Src} & {Attack} 
      & →Res50 & →Dense121 & →Res152 & →Inc-v3 & →Inc-v4 & →IncRes-v2 & →ViT & AVG \\
    \cmidrule{1-10}
 \multirow{13}{*}{\centering{\textbf{\textcolor{mybrown}{VGG19bn}}}}  & MIM & 0.56\% & 0.58\% & 0.29\% & 0.27\% & 0.20\% & 0.04\% & 0.07\% & 0.29\% \\
   & TIM & 3.80\% & 4.58\% & 1.84\% & 1.47\% & 1.09\% & 0.56\% & 0.16\% & 1.93\% \\
   & DIM & 2.98\% & 4.33\% & 1.76\% & 1.09\% & 1.16\% & 0.47\% & 0.16\% & 1.71\% \\
   & SI-NIM & 0.42\% & 0.42\% & 0.29\% & 0.11\% & 0.20\% & 0.18\% & 0.07\% & 0.24\% \\
   & Po-TI-Trip & 4.56\% & 6.27\% & 2.42\% & 1.69\% & 1.47\% & 0.87\% & 0.27\% & 2.51\% \\
&	$\text{SI-FGS}^2\text{M}$ & 4.12\% & 5.71\% & 1.97\% & 1.56\% & 1.25\% & 0.68\% & 0.25\% & 2.22\% \\
&   $\text{S}^2$I-SI-TI-DIM & 12.09\% & 14.84\% & 6.49\% & 4.87\% & 5.80\% & 2.93\% & 0.31\% & 6.76\% \\
&	Logit & 22.16\% & 30.47\% & 11.51\% & 5.51\% & 6.71\% & 1.91\% & 0.98\% & 11.32\% \\
&	DTMI-Logit-SU & 23.47\% & 31.47\% & 12.60\% & 6.13\% & 7.31\% & 2.07\% & 1.11\% & 12.02\%\\
&	RAP-LS & 24.31\% & 33.33\% & 13.56\% & 6.58\% & 8.51\% & 2.78\% & 1.11\% & 12.88\% \\
&	HGN & 34.18\% & 31.00\% & 19.36\% & 11.16\% & 8.87\% & 1.71\% & 2.26\% & 15.51\% \\
 &   TTP & 36.76\% & 34.44\% & 22.11\% & 11.82\% & 10.40\% & 2.20\% & 2.58\% & 17.19\% \\
   \rowcolor{gray!20} & ESMA (ours) & \underline{39.61}\% & \underline{47.25}\% & \underline{22.85}\% & \textbf{12.98}\% & \underline{11.73}\% & \textbf{4.09}\% & \underline{3.35}\% & \underline{20.27}\% \\
   \rowcolor{gray!20} & BEM-ESMA (ours) & \textbf{43.49}\% & \textbf{28.85}\% & \textbf{49.49}\% & \underline{12.73}\% & \textbf{15.31}\% & \underline{3.71}\% & \textbf{3.69}\% & \textbf{22.47}\% \\
    \midrule[0.8pt]
    \textbf{Src} & {Attack} 
      & →VGG19bn & →Res50 & →Dense121 & →Inc-v3 & →Inc-v4 & →IncRes-v2 & →ViT & AVG\\
    \cmidrule{1-10}
  \multirow{13}{*}{\centering{\textbf{\textcolor{mybrown}{Res152}}}} & MIM & 1.11\% & 5.69\% & 3.29\% & 0.69\% & 0.31\% & 0.18\% & 0.18\% & 1.64\% \\
   & TIM & 9.76\% & 26.64\% & 21.56\% & 5.69\% & 4.22\% & 3.04\% & 1.07\% & 10.28\% \\
   & DIM & 9.89\% & 26.27\% & 20.98\% & 5.69\% & 4.11\% & 2.44\% & 0.71\% & 10.01\% \\
   & SI-NIM & 0.82\% & 1.76\% & 1.16\% & 0.36\% & 0.20\% & 0.24\% & 0.11\% & 0.66\% \\
   & Po-TI-Trip & 14.53\% & 35.36\% & 29.87\% & 8.93\% & 6.29\% & 4.71\% & 1.36\% & 14.44\% \\
&	$\text{SI-FGS}^2\text{M}$ & 13.98\% & 39.22\% & 24.80\% & 7.98\% & 6.11\% & 4.31\% & 0.95\% & 13.91\% \\
&   $\text{S}^2$I-SI-TI-DIM & 19.29\% & 36.64\% & 30.78\% & 14.36\% & 11.42\% & 10.84\% & 1.29\% & 17.80\% \\
&	Logit & 35.44\% & 75.22\% & 61.31\% & 15.33\% & 10.69\% & 8.11\% & 2.76\% & 29.84\% \\
&	DTMI-Logit-SU & 37.76\% & 77.47\% & 62.91\% & 16.02\% & 11.51\% & 8.58\% & 3.02\% & 31.04\%\\
&	RAP-LS & 42.36\% & 83.20\% & 69.67\% & 18.82\% & 14.33\% & 10.36\% & 3.20\% & 34.56\% \\
&	HGN & 61.40\% & 73.31\% & 67.89\% & 33.47\% & 26.87\% & 11.31\% & 13.96\% & 41.17\% \\
 &   TTP & 65.31\% & 79.73\% & 74.93\% & 36.73\% & 30.11\% & 13.44\% & 15.62\% & 45.12\% \\
 \rowcolor{gray!20}  & ESMA (ours) & \underline{78.67}\% & \underline{88.18}\% & \underline{79.93}\% & \textbf{41.68}\% & \underline{34.38}\% & \textbf{14.58}\% & \underline{18.72}\% & \underline{50.88}\% \\
 \rowcolor{gray!20}  & BEM-ESMA (ours) & \textbf{80.81}\% & \textbf{89.76}\% & \textbf{81.57}\% & \underline{41.39}\% & \textbf{35.94}\% & \underline{14.11}\% & \textbf{19.33}\% & \textbf{51.84}\% \\
    \bottomrule[1pt]
    \end{tabular}
  \label{tab2}
\end{table*}

\paragraph{Victim Models}

We use four networks mentioned in \cite{21} as source models — ResNet-50 \cite{32} (Res50), VGG-19-bn \cite{33} (VGG19bn), DenseNet-121 \cite{34} (Dense121), ResNet-152 \cite{32} (Res152). In addition to these four models, we select three models from the Inception series: Inception-v3 \cite{35} (Inc-v3), Inception-v4 \cite{36} (Inc-v4), Inception-ResNet-v2 \cite{36} (IncRes-v2), and a transformer-based vision model, VIT \cite{49}. The analysis and fundamental assumptions of this paper are grounded in the condition of training data from the same distribution. To explore transferability between models trained on data with different distributions, we conduct additional transfer attacks on two adversarially trained models, Inc-v3-adv \cite{57}, and IncRes-v2-ens \cite{58}. Results are presented in Table \ref{tab2}. Furthermore, we assess the adversarial transferability of ESMA in scenarios where the source model is an ensemble of different models, and the corresponding results are detailed in Table \ref{tab4}.

\paragraph{Baseline Methods}

We have chosen numerous iterative instance-specific attack benchmarks, including MIM \cite{11}, SI-NIM \cite{12}, TIM \cite{13}, DIM \cite{15}, and advanced iterative instance-specific attacks that demonstrate competitiveness in target settings, such as Po-TI-Trip \cite{18}, Logit \cite{19}, Rap-LS \cite{51}, FGS$^2$M \cite{52}, DMTI-Logit-SU \cite{53}, S$^2$I-SI-TI-DIM \cite{54}. Our comparison also incorporates generative adversarial attacks, HGN \cite{55}, and TTP \cite{21}, where TTP stands as the current State of the Art (SOTA) generative method.

\paragraph{Parameter Settings} 

For the parameter settings of iterative attack methods, we adopt the default configurations from \cite{18}, with a total iteration number of $T=20$ and a step size of $\alpha=\epsilon/T$, where the $\ell_\infty$ perturbation restriction $\epsilon$ is fixed at $16/255$. Momentum factor $\mu$ is set to $1$. For the stochastic input diversity in DIM, the probability of applying input diversity is set to $0.7$. In the case of TIM, the kernel length is adjusted to $7$, which is more suitable for targeted attacks. Po-TI-Trip has a triplet loss weight $\lambda$ of $0.01$, and the margin $\gamma$ is set to $0.007$. Following \cite{19}, Logit is configured with $300$ iteration steps, and for RAP-LS, we choose $400$ iteration steps, setting $K_{LS}$ to $100$ and $\epsilon_n$ to $12/255$ \cite{51}. As for TTP, due to our use of a relatively small training set, we add $10$ epochs to the original paper's settings \cite{21} to ensure model performance. The learning rates for the Adam optimizer are $1e-4$ ($\beta_1=.5$, $\beta_2=.999$) for TTP and $2e-5$ for HGN, respectively. Finally, for our model, we employ the AdamW optimizer, training for $300$ epochs with a learning rate of $1e-4$ ($350$ for cases where the source model is VGG19bn, Dense121 Ensemble). The value of $q$ used for sample screening is set to $2$, and the $\nu$ of BEM is set to $1$.

\begin{figure*}[h!t]
\centering 
\subfigure[]{
\includegraphics[width=0.49\textwidth]{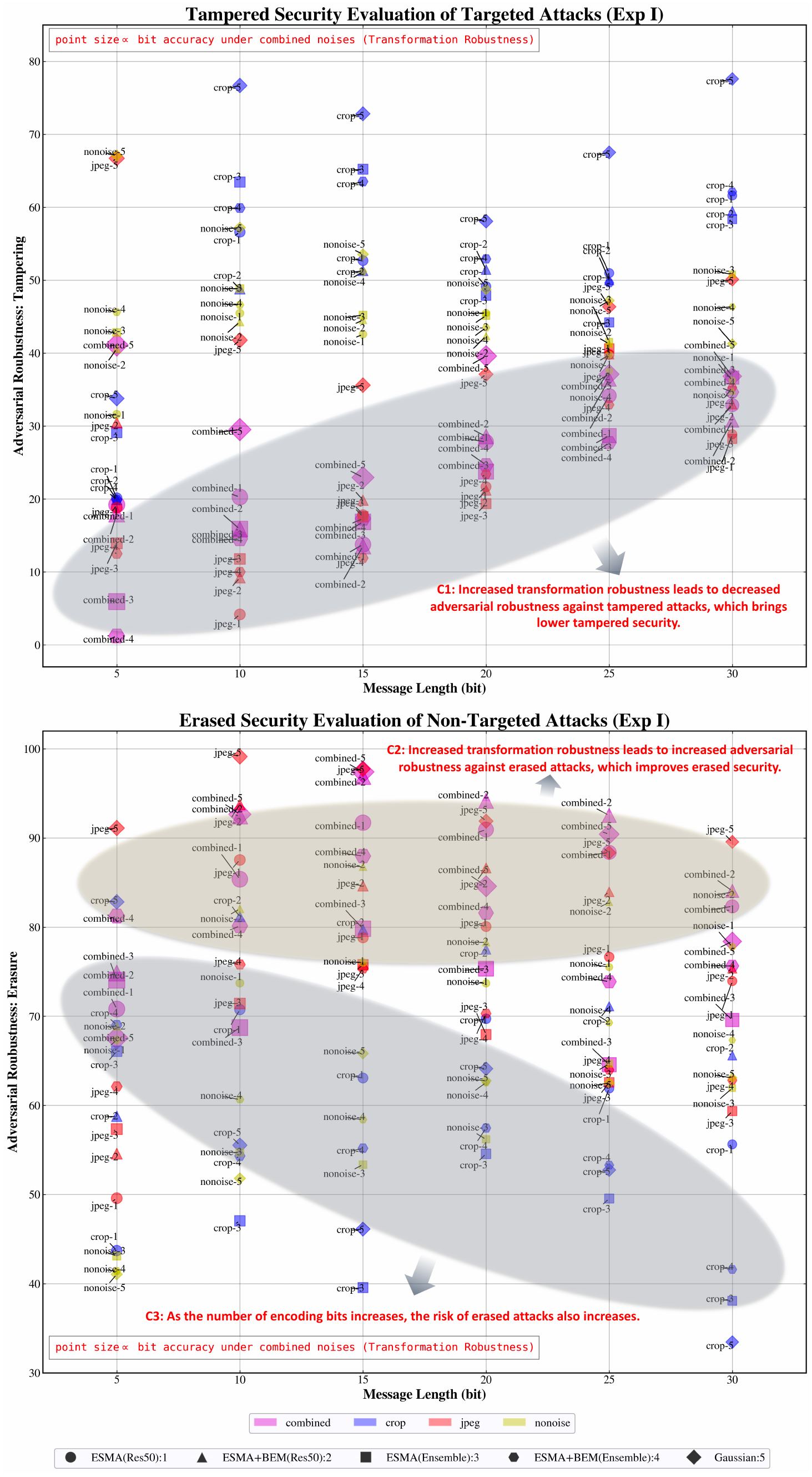}}
\subfigure[]{
\includegraphics[width=0.49\textwidth]{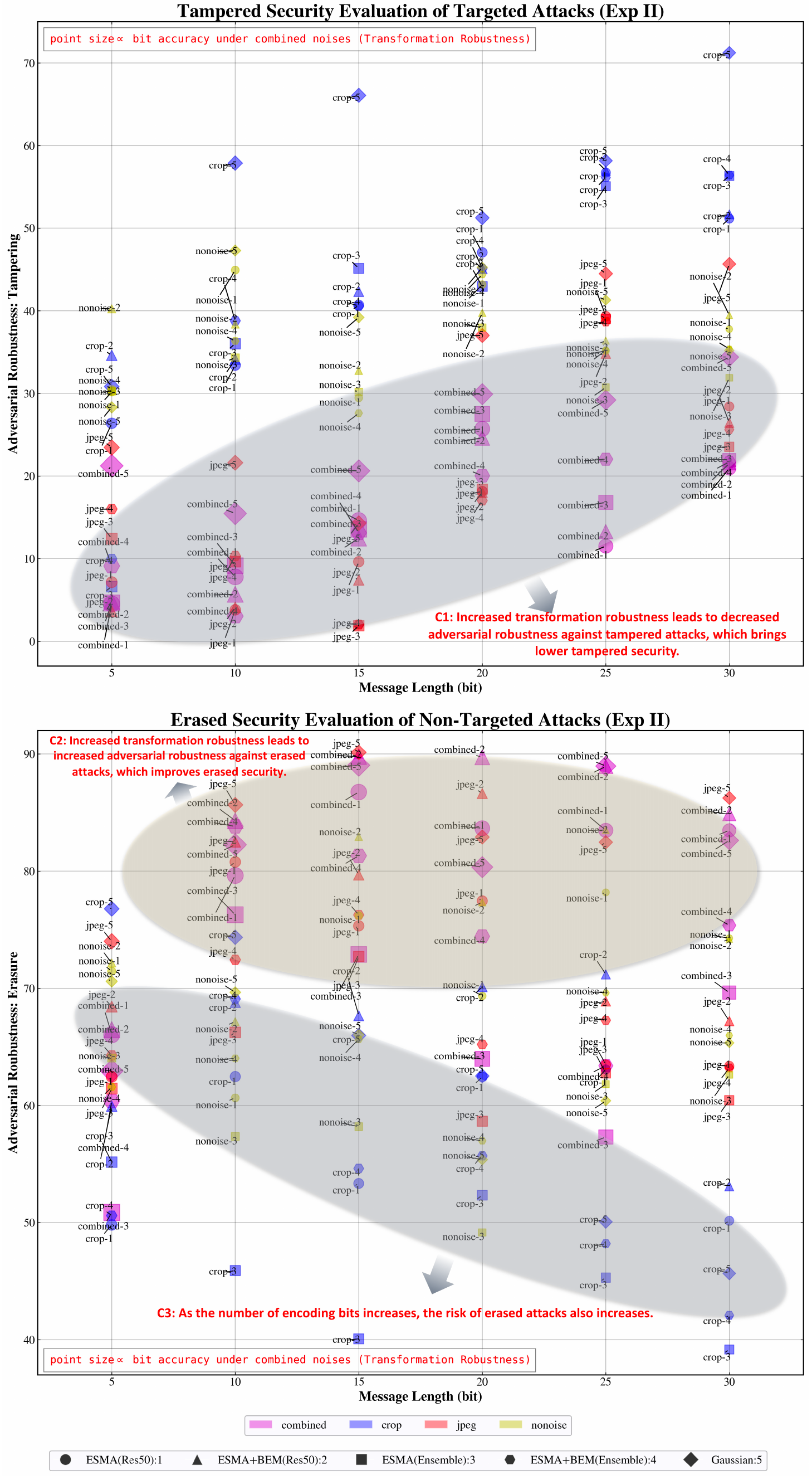}}
  \caption{
  (a): Targeted (tampering) and Non-targeted (erasure) attacks for Exp \Rmnum{1}. (b): Targeted (tampering) and Non-targeted (erasure) attacks for Exp \Rmnum{2}. The HiDDeN model trained under varying noise conditions, specifically involving combined-noise, crop, Jpeg, and nonoise are denoted by different colors. The results based on different src model (Res50/Ensemble) and different methods (ESMA/BEM-ESMA/Gaussian) are denoted by different markers. The x-axis is the message lenth. The y-axis is adversarial robustness for the certain attack. For tampering attack, the adversarial robustness is the bit error rate ($\mathcal E_{bit}^{tam}$). For erasure attack, the adversarial robustness is the bit correct rate ($1-\mathcal E_{bit}^{era}$). The point size is in proportion to the transformation robustness, which is the bit accuracy for different HiDDeN model under combined-transformation noises. }

  \label{fig 7}
\end{figure*}
\vspace{-3pt}

\paragraph{Evaluation Settings}

We train the generator on the training set of the selected ten classes, verifying targeted transferability on the validation set. For each target class, we use the $450$ images from the remaining classes as source data and perturb them to the target class. In the case of iterative instance-specific attacks, we directly target these instances and assess their success rates in targeted transfers. All methods, including training, were implemented on  1 $\times$ NVIDIA A6000 GPU.

\subsubsection{Targeted Attack Transferability Comparison}

Since our theoretical analysis is grounded in the assumption that the training data for the source and target models have the same distribution, we conducted experiments to investigate whether ESMA can achieve high transfer success rates between source and target models, regardless of whether they are trained on the same distribution or not. We evaluate and compare two experimental settings: one where both the source and target models are naturally trained (Table \ref{tab2}, \ref{tab3}), and another where the source model is naturally trained while the target model is adversarially trained (Table \ref{tab4}).

In our evaluation, ESMA's targeted attack success rate was tested for transfers from both a single model and an ensemble model to a naturally trained target model. ESMA demonstrated superior targeted transferability compared to other baseline methods, showing an average improvement of  $4\%-8\%$ in the success rate for targeted transfers. However, when subjected to additional tests on two adversarially trained models, Inc-v3-adv and IncRes-v2-ens, ESMA's performance did not maintain the same level of superiority observed in the previous experiments. This outcome aligns with our theoretical analysis, as our assumption was predicated on the source and target models being trained on the same distribution data. Therefore, \textbf{ESMA exhibited excellent adversarial transferability when targeting naturally trained models but did not outperform other baseline methods when applied to adversarially trained models}.

\begin{table}[htbp]
  \centering
  \caption{Targeted transfer success rates. The source model is an ensemble of Res-50, Inc-v3, and Inc-v4, the best results are bolded and the second are underlined.}
  \scriptsize
  \setlength{\tabcolsep}{1pt} 
  \renewcommand{\arraystretch}{0.55} 
    \begin{tabular}{c|ccccccccc}
    \toprule[0.7pt]
    \multirow{2}[2]{*}{Attack} & \multicolumn{6}{c}{\textbf{\textcolor{mybrown}{Src:Ensemble}}} \\
      & →VGG19bn & →Dense121 & →Res152  & →IncRes-v2 & →ViT & AVG\\
    \midrule
    MIM & 1.76\% & 3.40\% & 3.29\% & 1.07\% & 0.22\% & 1.95\% \\
    TIM  & 9.96\% & 16.13\% & 14.76\%  & 8.27\% & 1.11\% & 10.05\% \\
    DIM & 10.27\% & 16.31\% & 15.22\%  & 8.20\% & 0.87\% & 10.17\% \\
    SI-NIM  & 1.16\% & 1.67\% & 1.49\%  & 0.93\% & 0.13\% & 1.08\% \\
    Po-TI-Trip  & 14.22\% & 21.76\% & 18.69\%  & 12.09\% &  1.62\% & 13.68\% \\
   $\text{SI-FGS}^2\text{M}$ &  15.72\% & 24.84\% & 20.79\% & 12.20\% & 1.53\% & 15.02\% \\
	$\text{S}^2$I-SI-TI-DIM &  17.82\% & 21.07\% & 21.36\% & 16.38\% & 2.38\% & 15.80\% \\
	Logit   & 16.09\% & 27.56\% & 17.51\% & 36.96\% & 2.20\% & 20.06\% \\
	DTMI-Logit-SU & 18.11\% & 30.16\% & 18.73\% & 38.64\% & 2.64\% & 21.71\% \\
	RAP-LS    & 48.98\% & 72.91\% &  70.67\% & 26.53\% & 4.22\% & 44.66\% \\
	HGN   & 55.10\% & 67.37\% & 63.54\% & 49.85\% & 26.78\% & 52.53\% \\
	
    TTP  & 60.02\% & 71.33\% & 67.44\% & 55.42\% & 30.04\% & 56.85\% \\
    \rowcolor{gray!20}ESMA & \underline{62.51}\% & \underline{74.11}\% & \underline{72.61}\% &  \underline{62.33}\% &\underline{32.70}\% & \underline{60.85}\% \\
    \rowcolor{gray!20}BEM-ESMA & \textbf{63.65}\% & \textbf{76.27}\% & \textbf{73.83}\% &  \textbf{62.46}\% &\textbf{33.52}\% & \textbf{61.95}\% \\
    \bottomrule[0.8pt]
    \end{tabular}
  \label{tab3}
\end{table}
\vspace{-3pt}

\begin{table*}[t]
  \centering
  \caption{Targeted transfer success rates on adversarially trained models. "Src" indicates the source model, the best results are bolded and the second are underlined.}
  \tiny
  \renewcommand{\arraystretch}{1} 
    \begin{tabular}{c|c|cccccccccccccc}
    \toprule[1pt]
    \textbf{Src}& \textbf{Target}
      & MIM &TIM & DIM & SI-NIM &  Po-TI-Trip & $\text{SI-FGS}^2\text{M}$ & $\text{S}^2$I-SI-TI-DIM & Logit & DTMI-Logit-SU & RAP-LS & HGN & TTP & ESMA & BEM-ESMA\\
    \cmidrule{1-16}
   \multirow{2}{*}{\centering{\textbf{\textcolor{brown}{Res50}}}}& →$\text{Inc-v3}_\text{adv}$ & 0.11\% & 0.36\% & 0.18\% & 0.11\% & 0.22\% & 0.24\% & 0.76\% & 0.20\% & 0.27\% & 0.38\% & 2.12\% & \underline{3.53\%} & 3.23\% & \textbf{4.15\%}\\
   & →$\text{IncRes-v2}_\text{ens}$ & 0.11\% & 0.22\% & 0.18\% & 0.02\% & 0.38\% & 0.30\% & 0.89\% &  0.18\% & 0.22\% & 0.27\% & 2.70\% & \underline{4.53\%} & 3.98\% & \textbf{4.29\%} \\

    \midrule[0.8pt]
    \textbf{Src}& \textbf{Target}
      & MIM &TIM & DIM & SI-NIM &  Po-TI-Trip & $\text{SI-FGS}^2\text{M}$ & $\text{S}^2$I-SI-TI-DIM & Logit & DTMI-Logit-SU & RAP-LS & HGN & TTP & ESMA & BEM-ESMA\\
    \cmidrule{1-16}
   \multirow{2}{*}{\centering{\textbf{\textcolor{brown}{Dense121}}}}& →$\text{Inc-v3}_\text{adv}$ & 0.13\% & 0.13\% & 0.18\% & 0.11\% & 0.18\% & 0.22\% & 0.56\% & 0.20\% & 0.20\% & 0.22\% & 1.88\% & \underline{3.18\%} & 1.73\% & \textbf{3.55\%}\\
   & →$\text{IncRes-v2}_\text{ens}$ & 0.09\% & 0.18\% & 0.13\% & 0.11\% & 0.31\% & 0.20\% & 0.84\% & 0.22\% & 0.20\% & 0.18\% & 2.39\% & \underline{3.56\%} & 1.83\% & \textbf{3.94\%}\\

    \midrule[0.8pt]
    \textbf{Src}&\textbf{Target}
      & MIM &TIM & DIM & SI-NIM &  Po-TI-Trip & $\text{SI-FGS}^2\text{M}$ & $\text{S}^2$I-SI-TI-DIM & Logit & DTMI-Logit-SU & RAP-LS & HGN & TTP & ESMA & BEM-ESMA\\
    \cmidrule{1-16}
   \multirow{2}{*}{\centering{\textbf{\textcolor{brown}{VGG19bn}}}}& →$\text{Inc-v3}_\text{adv}$ & 0.13\% & 0.11\% & 0.16\% & 0.11\% & 0.11\% & 0.22\% & 0.16\% & 0.13\% & 0.13\% & 0.18\% & 0.20\% & \underline{0.49\%} & 0.31\% & \textbf{1.33\%}\\
   &→ $\text{IncRes-v2}_\text{ens}$ & 0.04\% & 0.07\% & 0.09\% & 0.01\% & 0.11\% & 0.13\% & 0.22\% & 0.07\% & 0.07\% & 0.09\% & 0.53\% & \underline{0.60\%} & 0.22\% & \textbf{0.89\%}\\

    \midrule[0.8pt]
    \textbf{Src}& \textbf{Target}
      & MIM &TIM & DIM & SI-NIM &  Po-TI-Trip & $\text{SI-FGS}^2\text{M}$ & $\text{S}^2$I-SI-TI-DIM & Logit & DTMI-Logit-SU & RAP-LS & HGN & TTP & ESMA & BEM-ESMA\\
    \cmidrule{1-16}
   \multirow{2}{*}{\centering{\textbf{\textcolor{brown}{Res152}}}}&→ $\text{Inc-v3}_\text{adv}$ & 0.13\% & 0.20\% & 0.20\% & 0.04\% & 0.40\% & 0.27\% & 0.93\% & 0.24\% & 0.26\% & 0.24\% & 4.26\% & \underline{5.96\%} & 2.31\% & \textbf{6.23\%}\\
   &→ $\text{IncRes-v2}_\text{ens}$ & 0.07\% & 0.36\% & 0.27\% & 0.07\% & 0.56\% & 0.33\% & 1.02\% & 0.27\% & 0.31\% & 0.38\% & 4.55\% & \underline{5.96\%} & 3.68\% & \textbf{7.04\%}\\

    \midrule[0.8pt]
    \textbf{Src}& \textbf{Target}
      & MIM &TIM & DIM & SI-NIM &  Po-TI-Trip & $\text{SI-FGS}^2\text{M}$ & $\text{S}^2$I-SI-TI-DIM & Logit & DTMI-Logit-SU & RAP-LS & HGN & TTP & ESMA & BEM-ESMA\\
    \cmidrule{1-16}
   \multirow{2}{*}{\centering{\textbf{\textcolor{brown}{Ensemble}}}}&→ $\text{Inc-v3}_\text{adv}$ & 0.20\% & 0.60\% & 0.40\% & 0.13\% & 0.69\% & 0.60\% & 2.51\% & 0.53\% & 0.53\% & 0.58\% & 22.67\% & \underline{24.44\%} & 17.00\% & \textbf{26.42\%}\\
   &→ $\text{IncRes-v2}_\text{ens}$ & 0.11\% & 0.96\% & 0.42\% & 0.09\% & 0.87\% & 0.66\% & 2.51\% & 0.20\% & 0.24\% & 0.58\% & 21.25\% & \underline{23.33\%} & 20.82\% & \textbf{25.31\%}\\

    \bottomrule[1pt]
    \end{tabular}
  \label{tab4}
\end{table*}

To assess the specific impact of the proposed BEM mechanism on ESMA's target attack transferability, we tested ESMA with BEM in two scenarios: when both the source and target models are naturally trained, and when the source model is naturally trained while the target model is adversarially trained. Given that the BEM mechanism diminishes ESMA's dependence on source data, \textbf{ESMA with the BEM mechanism not only elevates target attack transferability on naturally trained target models but also enhances transferability on adversarially trained target models}. Consequently, BEM-ESMA achieves higher target attack transfer success rates in both scenarios compared to all baseline methods.

\begin{figure}[h]
  \centering
  \includegraphics[width=0.4\textwidth]{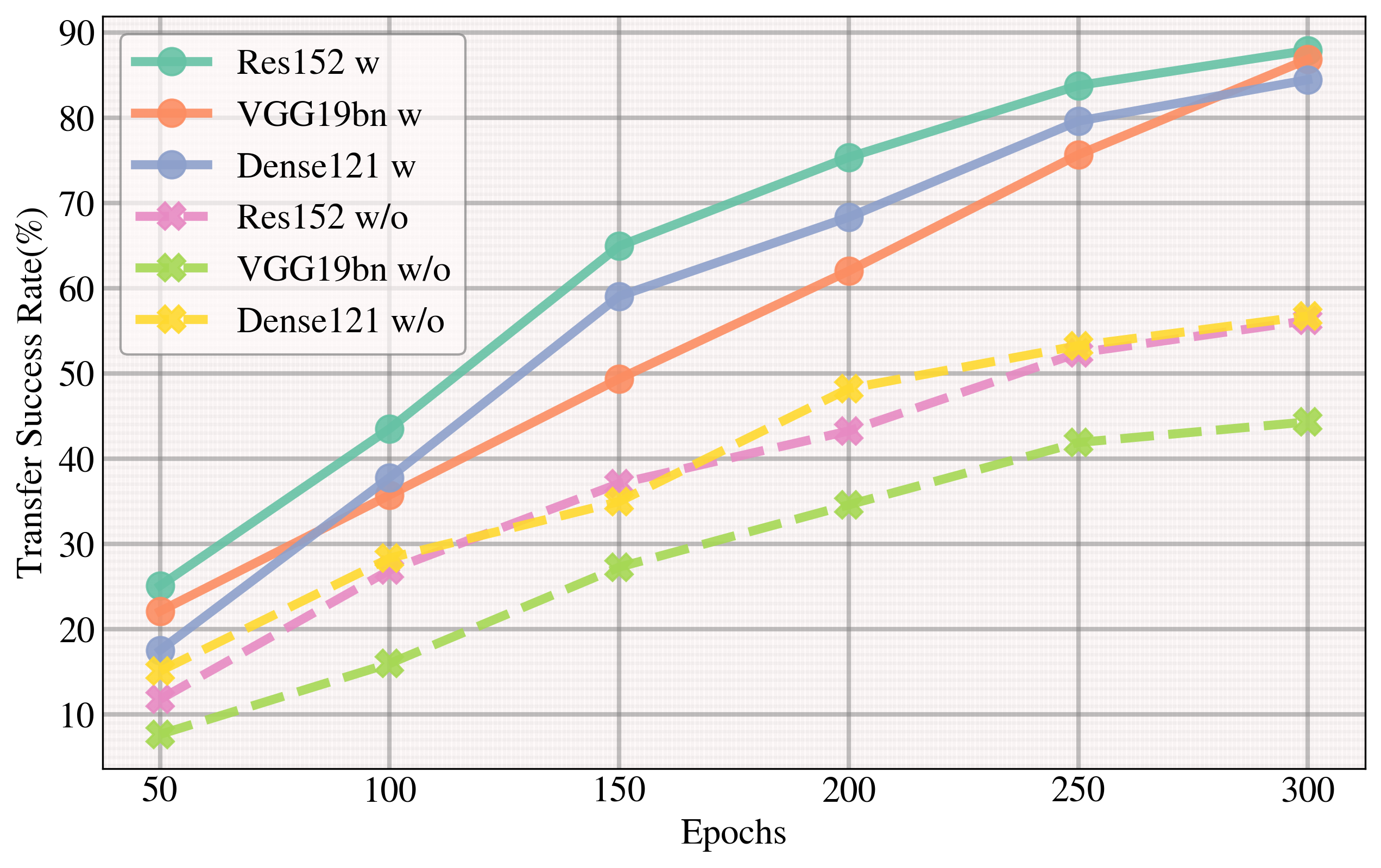} 
  \caption{Comparison of targeted attack transfer success rates with (w) pre-trained embeddings and without (w/o) pre-trained embeddings at different training epochs. Src:Res50.}
  \label{pre-train}
\end{figure}

\begin{table*}[t]
  \centering
  \scriptsize
  \setlength{\tabcolsep}{9.5pt}
  \renewcommand{\arraystretch}{0.25}
  \caption{Comparison of sample density (binned).}
    \begin{tabular}{c|c|cccccccccc}
    \toprule[1pt]
      & Counts & 0-0.1 & 0.1-0.2 & 0.2-0.3 & 0.3-0.4 & 0.4-0.5 & 0.5-0.6 & 0.6-0.7 & 0.7-0.8 & 0.8-0.9 & 0.9-1.0 \\
    \midrule[1pt]
    \multirow{2}[4]{*}{Case for $\epsilon=8/255$:} & Clean & 32 & 35 & 31 & 45 & 39 & 51 & 57 & 51 & 61 & 48 \\
\cmidrule{2-12}      & ESMA & 27 & 30 & 32 & 37 & 43 & 49 & 62 & 51 & 70 & 49 \\
    \midrule[1pt]
    \multirow{2}[4]{*}{Case for $\epsilon=16/255$:} & Clean & 32 & 35 & 31 & 45 & 39 & 51 & 57 & 51 & 61 & 48 \\
\cmidrule{2-12}      & ESMA & 25 & 27 & 36 & 30 & 49 & 50 & 59 & 54 & 71 & 49 \\
    \bottomrule[1pt]
    \end{tabular}
  \label{tab5}
\vspace{-5pt}
\end{table*}

\subsubsection{Ablation Study}

\textbf{Does the pre-training of class embeddings truly enhance ESMA's adversarial transferability?} To ascertain the efficacy of the pre-training embedding operation, we conducted a comparative analysis between scenarios with (w/) and without (w/o) pre-trained embeddings and target anchor screening. The results, depicted in Figure \ref{pre-train}, reveal a substantial performance gap when pre-trained embeddings are omitted, evident across various training durations. Furthermore, the inclusion of target anchor screening significantly enhances results compared to the random selection of target anchors.

\textbf{Does ESMA truly increase local sample density?} To validate ESMA's impact on boosting the sample density of the target class for original samples, we utilized samples from adversarial testing in Table \ref{tab2}. Initially, we computed the sample density of the target class for clean samples, denoted as $\rho_{(\text{target}, x_i, r)}$ (where $r$ is set to $600$). Subsequently, ESMA was applied to generate adversarial samples $x_i^{adv}$, and the sample density of the target class for these adversarial samples, denoted as $\rho_{(\text{target}, x_i^{adv}, r)}$, was calculated. In both instances, density calculation results were averaged across all target classes for each sample. We normalized the results by dividing them by the maximum value among all results in both cases. Additionally, we categorized and counted the number of test samples in various intervals under different perturbation constraints. The outcomes are detailed in Table \ref{tab5}. The target class sample density of samples perturbed using ESMA increased in bins with larger values, affirming ESMA's capability to enhance target class sample density.

\subsection{The risk of erasure and tampering evaluation of deep watermarking.}

Utilizing the proposed ESMA and BEM-ESMA, we subsequently devise a progressive series of task settings to simulate scenarios involving watermarked images from various sources for transfer attacks. This comprehensive evaluation aims to assess the vulnerabilities of current deep watermarking techniques against erasure and tampering risks through transfer attack strategies, considering both the same type and different types of watermarking models.

\paragraph{Tasks}
First, we introduce a series of settings for simulated tasks. Let's assume that the watermarked images obtained come from different companies, and users can only confirm the source of the image without any other auxiliary information. Typically, each company has an independent watermark model ${W_i}$ and a watermark message pool used for encoding, denoted as $MP_i = \left\{ {M_1},{M_2},..,,{M_k},...,{M_{{n_p}}}\right\}$, where ${M_k} \in {\left\{ {0,1} \right\}^L}$ is a binary steganographic code of length $L$. As the watermark models of different companies are independent, we assume that the watermark message pools of different enterprises are disjoint ($MP_i \ne MP_j, i \ne j$). In the normal watermark encoding process, each time, a message ${M_{in}} \in MP_i$ needs to be selected from the message pool and encoded into the cover image ${I_{co}}$ of shape $C \times H \times W$ to obtain the encoded image ${I_{en}}$. During the verification phase, the watermark model extracts the watermark message ${M_{out}}$ from ${I_{en}}$. If the extracted message satisfies ${M_{out}} \in MP_i$, it is considered that the image was produced by the company with ${W_i}$.

Our simulated task settings encompass a series of progressively evolving scenarios.

\noindent \textbf{Exp \Rmnum{1}}: 
The watermarking models of four companies employ the same watermark framework but undergo diverse training methods. Specifically, $W_1$ to $W_4$ are instantiated using HiDDeN trained with four distinct enhancement methods: no enhancement (nonoise), crop enhancement (crop), JPEG enhancement (jpeg), and a combination of crop and JPEG enhancement (combined). Additionally, each company's message pool contains only one piece of information ($n_p=1$).

Exp \Rmnum{1} represents a relatively ideal and simpler scenario in its setup. However, due to the restrictive assumption of having only one message in the message pool ($n_p=1$), we consider a more generalized scenario, namely Exp \Rmnum{2} below.

\noindent \textbf{Exp \Rmnum{2}}:
The watermark models of the companies are instantiated using the same method as in Exp \Rmnum{1}, where the four companies utilize the same watermark framework trained with different enhancement methods. However, in this case, the message pool size is increased to $n_p=8$. To simulate real-world scenarios where users have limited access to enterprise watermark models, each model $W_i$ randomly selects four out of eight messages from the message pool $MP_i = { {M_1},{M_2},..,{M_8}}$ for encoding into the cover image $I_{co}$.

Experiments Exp \Rmnum{1} and Exp \Rmnum{2} assume that each enterprise's $W_i$ is instantiated using the same watermark model architecture. However, various watermark model architectures exist, extending beyond HiDDeN. In addition to the conventional Vanilla encoder-decoder structure, alternatives include the Stable Signature based on Stable Diffusion and the Flow-based architecture FED. To more comprehensively evaluate the erasure and tampering risks associated with different watermark architectures, we introduce Exp \Rmnum{3}.

\noindent \textbf{Exp \Rmnum{3}}: 
Three companies utilize distinct watermark architectures, with $W_1$ to $W_3$ instantiated using HiDDeN, Stable Signature, and FED, respectively. The message pool design remains consistent with Exp \Rmnum{2}, featuring $n_p=8$. In the watermark encoding process, four messages are randomly selected from the message pool for each company.

To comprehensively explore the influence of $L$ selection on the watermark model's reliability, we considered all integer lengths within the range of $5$ to $30$, inclusive, across Experiments \Rmnum{1} to \Rmnum{3}.

\paragraph{Evaluation Metrics}
In the three aforementioned setups, Exp \Rmnum{1} to Exp \Rmnum{3}, to comprehensively evaluate the risks associated with erasing (non-targeted attack) and tampering (targeted attack) watermarks through transfer attacks, we carried out watermark erasure and watermark tampering assessments across all three configurations. Here, we will provide detailed explanations of our evaluation criteria.

Let $I_{en,i}^{\mathcal G}$ denote the generated adversarial image, where $I_{en,i}$ represents the image encoded with a watermark by enterprise $i$. In the watermark removal test, the attacker's objective is to maximize the disparity between the message $D_{W_{i}}\left(I_{en,i}^{\mathcal G}\right)$ decoded by the decoder $D_{W_{i}}$ of the trial model $W_i$ from the generated adversarial image $I_{en,i}^{\mathcal G}$ and the message $D_{W_{i}}\left({I_{en,i}}\right)$ decoded by $D_{W_{i}}$ from the original image ${I_{en,i}}$, i.e. $\text{Maximize }\left\|D_{W_{i}}\left({I_{en,i}}\right)-D_{W_{i}}\left(I_{en,i}^{\mathcal G}\right)\right\|$. If $D_{W_{i}}\left({I_{en,i}}\right)\ne D_{W_{i}}\left(I_{en,i}^{\mathcal G}\right)$, it is considered as a successful erasure of the watermark encoded by model $W_i$.
For the watermark tampering test, the objective is to minimize the difference between the message $D_{W_i}\left(I_{en,j}^{\mathcal G}\right)$ decoded by $D_{W_i}$ from the adversarial image $I_{en,j}^{\mathcal G}$ originating from another class $j\neq i$ and the messages in the message pool $MP_i$. This can be expressed as $\text{Minimize }\min_{M_k\in MP_i}\left\|D_{W_i}\left(I_{en,j}^{\mathcal G}\right)-M_k\right\|$. If $D_{W_i}\left(I_{en,j}^{\mathcal G}\right)\in MP_i$, it is considered a successful tampering of the watermark originally from enterprise $j$ to make it appear as if it originated from enterprise $i$.

To facilitate a more detailed evaluation, we introduce two primary metrics: the watermark bit error rate (indicating watermark dissimilarity) and the watermark detection rate (serving as a detection indicator).
For the message $M_{in,i}$ encoded by $W_i$ and its encoded $n$ images $\{I_{en,i}^j\}_{j=1}^n$, the bit error rate for erasure attacks $\mathcal E_{bit}^{era}$ is calculated as follows:
\begin{equation}
   \mathcal E_{bit}^{era} = \frac{1}{n}\sum_{j=1}^{n}\left\|D_{W_i}\left(I_{en,i}^j\right)-D_{W_i}\left(I_{en,i}^{\mathcal G, j}\right)\right\|/L,
\end{equation}
the detection rate $\mathcal E_{det}^{era}$ for erasure attacks can be calculated using the following equation:
\begin{equation}
\mathcal E_{det}^{era} = \frac{1}{n}\sum_{j=1}^n\mathbbm 1\left(D_{W_i}\left(I_{en,i}^j\right)\ne D_{W_i}\left(I_{en,i}^{\mathcal G, j}\right)\right).
\end{equation}
The two metrics mentioned above, $\mathcal E_{det}^{era}$ and $\mathcal E_{bit}^{era}$ are used to measure the risk of watermark erasure in transfer attacks. 

Similarly, for tampering attacks, considering a total of $N_e$ enterprises, the bit error rate $\mathcal E_{bit}^{tam}$ can be calculated as follows:
\begin{equation}
\mathcal E_{bit}^{tam} =\frac{1}{n(N_e-1)}\sum_{j\ne i}\sum_{k=1}^{n}\max_{M_k\in MP_i}\frac{\left\|M_k-D_{W_i}\left(I_{en,j}^{\mathcal G, k}\right)\right\|}{L},
\end{equation}
likewise, the detection rate $\mathcal E_{det}^{tam}$ for tampering attacks can be calculated using the following equation:
\begin{equation}
\mathcal E_{det}^{era} = \frac{1}{n(N_e-1)}\sum_{j\ne i}\sum_{k=1}^n\max_{M_k\in MP_i}\mathbbm 1\left(M_k= D_{W_i}\left(I_{en,j}^{\mathcal G, k}\right)\right).
\end{equation}
The two metrics, $\mathcal E_{det}^{tam}$ and $\mathcal E_{bit}^{tam}$, mentioned above measure the risk of watermark leakage encoded by the watermark model $W_i$. They also assess the risk of the watermark verification mechanism being compromised.

Furthermore, regarding the feasibility of transfer attacks on watermark models, we consider two scenarios to speculate about their causes: 1) Watermarks generated by the watermark model can indeed be learned as a certain pattern by surrogate models (leakage occurs), enabling the effectiveness of transfer attacks. 2) The watermarks generated by the watermark model are inherently sensitive to noise, allowing even random noise to execute black-box attacks. To seek answers to these questions, we have additionally included a baseline of black-box attacks using Gaussian random noise as a reference, aiding in determining whether the effectiveness of transfer attacks stems from reason 1) or 2).

\subsubsection{Implementation Details}
\paragraph{Watermarking Baselines}

We select HiDDeN\cite{66}, Stable Signature\cite{75}, and FED\cite{78} as the three evaluated watermarking baselines. HiDDeN's training dataset and settings align with those in \cite{66}, randomly selecting $10,000$ images from MS COCO\cite{79} for training and $1,000$ for validation. To conduct robustness evaluations at various bit lengths, we choose a series of encoding lengths $L$ for HiDDeN. The image size is uniformly scaled to $H=128$, $W=128$ in Exp \Rmnum{1} and Exp \Rmnum{2}. For Exp \Rmnum{3},  Stable Signature's training follows the settings in \cite{75}, involving the training of eight fine-tuned Stable Diffusion decoders corresponding to eight message keys at each message length $L$. 
For FED, certain parameters such as the dataset and training parameters are kept consistent with \cite{78}. Given the distinctive structure of the FED model, its message length $L$ must be a square of an integer. Consequently, we continued training the FED model with a 64-bit message length $L$. However, in order to align with the various message lengths $L$ in Exp \Rmnum{3}, we opted to discard information beyond $L$ for each experiment. To align with Exp \Rmnum{1} and Exp \Rmnum{2}, the image size in Exp \Rmnum{3} of Stabel signature and FED is also $128 \times 128$.

\paragraph{Datasets for Src Model Training and Testing}

To create datasets with different watermarks, we randomly select $5000$ images from the MS COCO test set as the original images ${I_{co}}$ for the ESMA training set and $1000$ for the related test set. For each dataset creation, we apply watermark encoding refer to the respective ${W_i}$, $L$, $M{P_i}$, and ${n_p}$ to the aforementioned images to obtain the watermarked datasets consist of $I_{en}$ for training and testing the substitute model.

\paragraph{ESMA and BEM-ESMA Settings}

Watermark attack strategy hinges on employing a classifier trained on steganographic images from various sources as a surrogate model for executing transfer attacks. This classifier categorizes data from different enterprises, learning distinctive watermark message features for each enterprise. To fulfill this purpose, we use a pre-trained ResNet-50\cite{36} (Res50) with early-stopping in classification training, serving as the surrogate model for transfer attacks.

During ESMA training, we enhance the visual quality of the perturbated image $I^{\mathcal G}_{en}$ by replacing the original perturbation radius constraint with the introduction of an \textit{image distortion loss}, i.e., the $l_{2}$ distance between $I_{en}$ and ${I^{\mathcal G}}$ in the loss function:
\begin{equation}
\mathcal{L}_{I}\left\{ {{I_{en}}, I_{en}^{\mathcal G}} \right\} = \left\| {{I_{en}} - I_{en}^{\mathcal G}} \right\|^2/(CHW).
\end{equation}
Through grid search, we set the coefficient to $150$ to achieve a balance, where the quality of images generated by both ESMA and BEM-ESMA matches that of deep watermarking models (i.e. the difference between the watermarked images and the perturbated images is extremely small), while maximizing its adversarial attack capability.

\begin{table*}[t]
\scriptsize
  \centering
  \setlength{\tabcolsep}{4pt} 
  \renewcommand{\arraystretch}{0.5} 
  \caption{Attacking results for Exp \Rmnum{3} ($\%$). For tampering attack, the $/$ datas reprensents the detection rate $\mathcal E_{det}^{tam}$ / the bit correct rate ($1-\mathcal E_{bit}^{tam}$). For erasure attack, the $/$ datas reprensents the detection failure rate ($1 - \mathcal E_{det}^{era}$) / the bit error rate $\mathcal E_{bit}^{era}$. }
\begin{tabular}{@{}ccccllcccc@{}}
\toprule
\multicolumn{1}{c|}{Victim Model} & \multicolumn{1}{c|}{Attack Type} & \multicolumn{1}{c|}{Src} & \multicolumn{1}{c|}{Method} & \multicolumn{1}{c}{5} & \multicolumn{1}{c}{10} & 15 & 20 & 25 & 30 \\ \midrule
\multirow{10}{*}{FED} & \multirow{5}{*}{Tampering} & - & Gaussian & \multicolumn{1}{c}{0.67/79.52} & \multicolumn{1}{c}{0.46/72.43} & \multicolumn{1}{c}{0.00/68.99} & \multicolumn{1}{c}{0.00/66.06} & \multicolumn{1}{c}{0.00/64.62} & \multicolumn{1}{c}{0.00/63.21} \\ \cmidrule(l){3-10} 
 &  & \multirow{2}{*}{Res50} & ESMA & \multicolumn{1}{c}{30.71/85.17} & \multicolumn{1}{c}{0.09/73.89} & \multicolumn{1}{c}{0.00/68.98} & \multicolumn{1}{c}{0.00/66.84} & \multicolumn{1}{c}{0.00/65.25} & \multicolumn{1}{c}{0.00/63.09} \\
 &  &  & BEM-ESMA & \multicolumn{1}{c}{24.64/81.65} & \multicolumn{1}{c}{14.29/74.63} & \multicolumn{1}{c}{0.01/68.65} & \multicolumn{1}{c}{0.00/66.88} & \multicolumn{1}{c}{0.00/65.92} & \multicolumn{1}{c}{0.00/63.66} \\ \cmidrule(l){3-10} 
 &  & \multirow{2}{*}{Ensemble} & ESMA & \multicolumn{1}{c}{21.79/81.01} & \multicolumn{1}{c}{0.51/72.70} & \multicolumn{1}{c}{0.01/70.80} & \multicolumn{1}{c}{0.00/67.57} & \multicolumn{1}{c}{0.00/64.04} & \multicolumn{1}{c}{0.00/63.30} \\
 &  &  & BEM-ESMA & \multicolumn{1}{c}{23.21/81.76} & \multicolumn{1}{c}{0.92/72.09} & \multicolumn{1}{c}{0.00/68.99} & \multicolumn{1}{c}{0.00/66.74} & \multicolumn{1}{c}{0.00/64.36} & \multicolumn{1}{c}{0.00/63.36} \\ \cmidrule(l){2-10} 
 & \multirow{5}{*}{Erasure} & - & Gaussian & \multicolumn{1}{c}{75.15/35.13} & \multicolumn{1}{c}{91.28/38.25} & \multicolumn{1}{c}{94.05/40.43} & \multicolumn{1}{c}{97.13/40.40} & \multicolumn{1}{c}{97.43/42.19} & \multicolumn{1}{c}{98.19/41.11} \\ \cmidrule(l){3-10} 
 &  & \multirow{2}{*}{Res50} & ESMA & \multicolumn{1}{c}{74.14/34.35} & \multicolumn{1}{c}{92.84/37.63} & \multicolumn{1}{c}{94.51/40.76} & \multicolumn{1}{c}{96.82/41.30} & \multicolumn{1}{c}{97.28/40.34} & \multicolumn{1}{c}{96.15/39.55} \\
 &  &  & BEM-ESMA & \multicolumn{1}{c}{70.21/33.42} & \multicolumn{1}{c}{87.40/36.49} & \multicolumn{1}{c}{92.04/39.73} & \multicolumn{1}{c}{95.01/40.14} & \multicolumn{1}{c}{93.20/41.22} & \multicolumn{1}{c}{96.90/40.54} \\ \cmidrule(l){3-10} 
 &  & \multirow{2}{*}{Ensemble} & ESMA & \multicolumn{1}{c}{81.65/37.80} & \multicolumn{1}{c}{91.99/38.83} & \multicolumn{1}{c}{96.52/44.21} & \multicolumn{1}{c}{93.55/40.25} & \multicolumn{1}{c}{97.88/42.48} & \multicolumn{1}{c}{98.29/40.75} \\
 &  &  & BEM-ESMA & \multicolumn{1}{c}{77.82/35.21} & \multicolumn{1}{c}{90.02/37.76} & \multicolumn{1}{c}{93.75/41.94} & \multicolumn{1}{c}{93.60/40.26} & \multicolumn{1}{c}{95.82/41.89} & \multicolumn{1}{c}{96.77/40.22} \\ \midrule
\multirow{10}{*}{HiDDeN} & \multirow{5}{*}{Tampering} & - & Gaussian & \multicolumn{1}{c}{0.52/79.76} & \multicolumn{1}{c}{0.00/64.82} & \multicolumn{1}{c}{0.00/74.12} & \multicolumn{1}{c}{0.00/66.10} & \multicolumn{1}{c}{0.00/63.65} & \multicolumn{1}{c}{0.00/62.36} \\ \cmidrule(l){3-10} 
 &  & \multirow{2}{*}{Res50} & ESMA & \multicolumn{1}{c}{30.37/85.91} & \multicolumn{1}{c}{0.00/69.01} & \multicolumn{1}{c}{0.00/70.35} & \multicolumn{1}{c}{0.00/68.26} & \multicolumn{1}{c}{0.00/64.22} & \multicolumn{1}{c}{0.00/64.32} \\
 &  &  & BEM-ESMA & \multicolumn{1}{c}{24.80/84.81} & \multicolumn{1}{c}{0.00/65.99} & \multicolumn{1}{c}{0.05/74.02} & \multicolumn{1}{c}{0.00/71.34} & \multicolumn{1}{c}{0.00/62.55} & \multicolumn{1}{c}{0.00/64.90} \\ \cmidrule(l){3-10} 
 &  & \multirow{2}{*}{Ensemble} & ESMA & \multicolumn{1}{c}{32.53/84.70} & \multicolumn{1}{c}{0.00/65.16} & \multicolumn{1}{c}{0.05/71.39} & \multicolumn{1}{c}{0.00/64.01} & \multicolumn{1}{c}{0.00/61.29} & \multicolumn{1}{c}{0.00/62.26} \\
 &  &  & BEM-ESMA & \multicolumn{1}{c}{12.95/80.39} & \multicolumn{1}{c}{0.05/68.29} & \multicolumn{1}{c}{0.00/73.83} & \multicolumn{1}{c}{0.00/67.02} & \multicolumn{1}{c}{0.00/62.87} & \multicolumn{1}{c}{0.00/62.91} \\ \cmidrule(l){2-10} 
 & \multirow{5}{*}{Erasure} & - & Gaussian & \multicolumn{1}{c}{93.65/33.28} & \multicolumn{1}{c}{99.58/47.77} & \multicolumn{1}{c}{95.73/36.56} & \multicolumn{1}{c}{99.90/43.37} & \multicolumn{1}{c}{98.80/42.90} & \multicolumn{1}{c}{100.00/42.52} \\ \cmidrule(l){3-10} 
 &  & \multirow{2}{*}{Res50} & ESMA & \multicolumn{1}{c}{90.16/45.21} & \multicolumn{1}{c}{100.00/51.58} & \multicolumn{1}{c}{99.53/45.18} & \multicolumn{1}{c}{100.00/45.27} & \multicolumn{1}{c}{100.00/47.81} & \multicolumn{1}{c}{100.00/47.81} \\
 &  &  & BEM-ESMA & \multicolumn{1}{c}{88.54/43.15} & \multicolumn{1}{c}{99.84/49.51} & \multicolumn{1}{c}{98.65/41.74} & \multicolumn{1}{c}{100.00/42.45} & \multicolumn{1}{c}{99.95/45.36} & \multicolumn{1}{c}{100.00/43.59} \\ \cmidrule(l){3-10} 
 &  & \multirow{2}{*}{Ensemble} & ESMA & \multicolumn{1}{c}{95.73/47.58} & \multicolumn{1}{c}{100.00/52.41} & \multicolumn{1}{c}{100.00/48.84} & \multicolumn{1}{c}{100.00/48.29} & \multicolumn{1}{c}{100.00/46.43} & \multicolumn{1}{c}{100.00/50.04} \\
 &  &  & BEM-ESMA & \multicolumn{1}{c}{96.30/51.64} & \multicolumn{1}{c}{99.38/51.19} & \multicolumn{1}{c}{100.00/51.10} & \multicolumn{1}{c}{99.95/44.90} & \multicolumn{1}{c}{100.00/46.18} & \multicolumn{1}{c}{100.00/48.48} \\ \midrule
\multirow{10}{*}{Stable   Signature} & \multirow{5}{*}{Tampering } & - & Gaussian & \multicolumn{1}{c}{0.58/80.02} & \multicolumn{1}{c}{0.00/70.52} & \multicolumn{1}{c}{0.00/65.73} & \multicolumn{1}{c}{0.00/59.80} & \multicolumn{1}{c}{0.00/66.74} & \multicolumn{1}{c}{0.00/64.18} \\ \cmidrule(l){3-10} 
 &  & \multirow{2}{*}{Res50} & ESMA & \multicolumn{1}{c}{18.85/81.65} & \multicolumn{1}{c}{0.41/73.57} & \multicolumn{1}{c}{0.00/69.09} & \multicolumn{1}{c}{0.00/64.49} & \multicolumn{1}{c}{0.00/68.58} & \multicolumn{1}{c}{0.00/61.89} \\
 &  &  & BEM-ESMA & \multicolumn{1}{c}{9.38/76.43} & \multicolumn{1}{c}{0.56/73.88} & \multicolumn{1}{c}{0.00/72.67} & \multicolumn{1}{c}{0.00/63.57} & \multicolumn{1}{c}{0.00/68.71} & \multicolumn{1}{c}{0.00/64.50} \\ \cmidrule(l){3-10} 
 &  & \multirow{2}{*}{Ensemble} & ESMA & \multicolumn{1}{c}{8.61/80.52} & \multicolumn{1}{c}{0.15/74.02} & \multicolumn{1}{c}{0.00/65.69} & \multicolumn{1}{c}{0.00/65.82} & \multicolumn{1}{c}{0.00/61.36} & \multicolumn{1}{c}{0.00/62.44} \\
 &  &  & BEM-ESMA & \multicolumn{1}{c}{5.69/78.91} & \multicolumn{1}{c}{0.10/73.16} & \multicolumn{1}{c}{0.00/69.84} & \multicolumn{1}{c}{0.00/64.31} & \multicolumn{1}{c}{0.00/62.76} & \multicolumn{1}{c}{0.00/61.04} \\ \cmidrule(l){2-10} 
 & \multirow{5}{*}{Erasure} & - & Gaussian & \multicolumn{1}{c}{60.05/29.80} & \multicolumn{1}{c}{83.30/24.18} & \multicolumn{1}{c}{99.30/28.39} & \multicolumn{1}{c}{100.00/36.45} & \multicolumn{1}{c}{89.20/9.90} & \multicolumn{1}{c}{99.90/19.30} \\ \cmidrule(l){3-10} 
 &  & \multirow{2}{*}{Res50} & ESMA & \multicolumn{1}{c}{59.80/20.57} & \multicolumn{1}{c}{77.95/20.16} & \multicolumn{1}{c}{91.70/20.11} & \multicolumn{1}{c}{100.00/40.46} & \multicolumn{1}{c}{99.85/19.72} & \multicolumn{1}{c}{99.85/19.72} \\
 &  &  & BEM-ESMA & \multicolumn{1}{c}{51.00/16.46} & \multicolumn{1}{c}{67.65/16.41} & \multicolumn{1}{c}{77.25/10.66} & \multicolumn{1}{c}{100.00/27.63} & \multicolumn{1}{c}{85.00/10.36} & \multicolumn{1}{c}{98.30/14.12} \\ \cmidrule(l){3-10} 
 &  & \multirow{2}{*}{Ensemble} & ESMA & \multicolumn{1}{c}{82.40/35.67} & \multicolumn{1}{c}{85.75/26.89} & \multicolumn{1}{c}{100.00/38.26} & \multicolumn{1}{c}{100.00/33.13} & \multicolumn{1}{c}{99.65/21.96} & \multicolumn{1}{c}{100.00/30.31} \\
 &  &  & BEM-ESMA & \multicolumn{1}{c}{55.70/21.12} & \multicolumn{1}{c}{68.50/17.01} & \multicolumn{1}{c}{99.75/31.55} & \multicolumn{1}{c}{99.90/25.95} & \multicolumn{1}{c}{94.90/12.47} & \multicolumn{1}{c}{99.40/22.25} \\ \bottomrule
\end{tabular}
\label{tab case 3}
\end{table*}

\begin{table}[t]
\centering
  \setlength{\tabcolsep}{1.5pt} 
  \renewcommand{\arraystretch}{1} 
  \caption{PSNR during watermarking and attacking process. For Exp \Rmnum{1} and Exp \Rmnum{2}, the watermark PSNR is the average of different HiDDeN model, while for Exp \Rmnum{3}, the watermark PSNR is the average of FED, HiDDeN and Stable Signature.}
\begin{tabular}{@{}ccccccccc@{}}
\toprule
Experiments & Src Model & Method & 5 & 10 & 15 & 20 & 25 & 30 \\ \midrule
\multirow{6}{*}{Exp \Rmnum{1}} & - & Watermark & 29.87 & 32.46 & 32.55 & 33.06 & 32.74 & 32.42 \\ \cmidrule(l){2-9} 
 & - & Gaussian & 27.52 & 27.51 & 27.51 & 27.51 & 27.51 & 27.50 \\ \cmidrule(l){2-9} 
 & \multirow{2}{*}{Res50} & ESMA & 29.16 & 28.83 & 29.72 & 30.03 & 30.88 & 31.10 \\
 &  & BEM-ESMA & 29.63 & 30.68 & 31.10 & 30.96 & 32.15 & 32.41 \\ \cmidrule(l){2-9} 
 & \multirow{2}{*}{Ensemble} & ESMA & 24.31 & 25.48 & 24.98 & 24.26 & 24.81 & 25.24 \\
 &  & BEM-ESMA & 25.90 & 26.33 & 27.17 & 25.45 & 27.74 & 26.75 \\ \midrule
\multirow{6}{*}{Exp \Rmnum{2}} & - & Watermark & 29.56 & 31.98 & 32.42 & 32.79 & 32.56 & 32.42 \\ \cmidrule(l){2-9} 
 & - & Gaussian & 27.51 & 27.52 & 27.51 & 27.51 & 27.5 & 27.50 \\ \cmidrule(l){2-9} 
 & \multirow{2}{*}{Res50} & ESMA & 29.45 & 29.32 & 29.16 & 29.87 & 29.82 & 29.75 \\
 &  & BEM-ESMA & 30.81 & 30.42 & 31.14 & 31.59 & 31.39 & 31.32 \\ \cmidrule(l){2-9} 
 & \multirow{2}{*}{Ensemble} & ESMA & 25.25 & 26.42 & 25.52 & 23.95 & 24.02 & 25.68 \\
 &  & BEM-ESMA & 27.21 & 28.94 & 27.95 & 25.36 & 25.68 & 27.26 \\ \midrule
\multirow{6}{*}{Exp \Rmnum{3}} & - & Watermark & 35.14 & 36.34 & 36.37 & 36.54 & 36.43 & 36.23 \\ \cmidrule(l){2-9} 
 & - & Gaussian & 27.48 & 27.50 & 27.50 & 27.48 & 27.49 & 27.49 \\ \cmidrule(l){2-9} 
 & \multirow{2}{*}{Res50} & ESMA & 28.51 & 27.00 & 27.65 & 27.22 & 27.94 & 28.66 \\
 &  & BEM-ESMA & 29.22 & 28.56 & 28.73 & 28.27 & 29.45 & 29.78 \\ \cmidrule(l){2-9} 
 & \multirow{2}{*}{Ensemble} & ESMA & 26.98 & 26.70 & 22.05 & 29.03 & 26.81 & 25.57 \\
 &  & BEM-ESMA & 25.75 & 28.28 & 25.70 & 30.15 & 27.86 & 28.58 \\ \bottomrule
\end{tabular}
\label{tab psnr}
\end{table}

\subsubsection{Risk assessment of watermark erasure and tampering.}
\textbf{Q1: What is the extent of the risk of deep watermarking being subjected to non-targeted attacks and targeted attacks for erasure or tampering?}
In order to fully evaluate the risk of deep watermarking being subjected to erasure or tampering through non-targeted and targeted transfer attacks, we first conduct non-targeted and transfer attacks testing on HiDDeN with different enhancement methods at various coding lengths according to the settings of Exp \Rmnum{1} and \Rmnum{2}, results are shown in Figure \ref{fig 7}. This was done to assess the erasure and tampering risks of HiDDeN trained with different enhancement methods under these two settings. The results indicate that regardless of the type of data augmentation, it cannot effectively resist both watermark erasure and tampering simultaneously. Specifically, when $L = 5$, under the condition that the source model is Ensemble, BEM-ESMA in Exp \Rmnum{1} can tamper over $90\%$ of the watermarks to HiDDeN watermarks trained with combined enhancement, and BEM-ESMA in Exp \Rmnum{2} can erase nearly $100\%$ of the watermarks to HiDDeN watermarks trained with JPEG enhancement. Moreover, regardless of whether it is in the Exp \Rmnum{1} or \Rmnum{2} setting, BEM-ESMA using the Ensemble source model almost guarantees an erasure rate $\mathcal E^{era}_{det}$ of over $80\%$ for HiDDeN under all enhancement methods. Although the success rate of watermark tampering decreases as $L$ increases, in Exp \Rmnum{1} and \Rmnum{2}, the risk of watermark erasure using ESMA and BEM-ESMA actually increases from an initial minimum of at least $60\%$ to nearly $100\%$ on HiDDeN under all enhancement methods. Therefore, a shorter message length brings a greater risk of tampering, while a longer message encoding length brings a greater risk of erasure. This means that changing the encoding length does not simultaneously reduce the risks of watermark erasure and tampering. Specifically, when the message length is $10$, using BEM-ESMA with Res50 as the source model can tamper $14.29\%$ of other source watermarks to appear as Stable Signature watermarks. When the message length is $5$, using Res50 as the source model, ESMA can achieve success rates of $30.71\%$ and $18.85\%$ in tampering other source watermarks to be validated as Stable Signature and FED watermarks, respectively. Similarly to Exp \Rmnum{1} and \Rmnum{2}, as the message length $L$ increases, the erasure success rate $\mathcal E^{era}_{det}$ for all watermark models rises to nearly $100\%$. These findings suggest that even the state-of-the-art watermark models currently available are associated with high risks of watermark erasure and tampering, and longer encoding lengths do not effectively reduce these risks. Additionally, we calculated the average PSNR before and after various erasure and tampering operations using different methods, and the results are shown in Table \ref{tab psnr}. Our erasure and tampering operations can achieve comparable PSNR values to that between original watermarked image and cover image, allowing for invisivle changes.

\textbf{Q2: For a single watermark model, what is the relationship between the robustness in image transformation and the adversarial robustness?} 
Based on the watermark erasure and tampering experiments conducted on various HiDDeN models trained with different data augmentation methods, we observed that as the robustness of HiDDeN to image transformations increases, the likelihood of its generated watermarks being erased decreases, as shown in the erasure attack in Figure \ref{fig 7}.

However, the likelihood of other classes being tampered with to resemble the target class through targeted attacks increases. This is because more powerful watermarks designed to withstand image transformations are more robust against minor perturbations, but they are also more prone to exhibiting fixed patterns. Although they are harder to erase, proxy models can more easily learn these fixed patterns, providing more guidance for targeted attacks. To illustrate this conclusion more clearly, we set a baseline by perturbing the watermark with Gaussian noise of the same magnitude as adversarial noise. The experiments showed that HiDDeN trained with data augmentation was more susceptible to being targeted for tampering compared to HiDDeN without augmentation (nonoise). Conversely, HiDDeN without augmentation, which had poor robustness against image transformations, had lower success rates when being target for tampering. This indicates that for more robust watermarks, they are more likely to be learned as patterns by proxy models, thus providing richer information guidance for attacks. This means that for individual models, even though we enhance the robustness of watermarks to transformations through various data augmentation techniques, it actually increases their risk of being target for tampering.

\textbf{Q3: If the watermarks come from encoding models with different structures, does the constraint between transformation robustness and adversarial robustness still hold?} 
The transfer attack tests conducted on different watermark architectures under the Exp \Rmnum{3} (as shown in Table \ref{tab case 3}) setting show that the watermark produced by FED, which has the most robust transformation, has the lowest probability of being erased by ESMA. Specifically, when $L\ge 20$, its erasure success rate is significantly lower compared to Stable Signature and HiDDeN. However, it is the easiest to tamper with other source watermarks and convert them into watermarks originating from FED using ESMA. Particularly, when $L=10$, BEM-ESMA with the Ensemble source model achieves a success rate of $14.29\%$ in tampering other source watermarks to appear as if they originated from FED. This suggests that in different model architectures, stronger watermark transformation robustness still makes it more susceptible to being target for tampering.

\section{Conclusion and Discussion}
Understanding the security of imperceptible deep watermarks is crucial, yet it remains a seriously unexplored problem. 
We undertake, for the first time, an evaluation and analysis from the perspective of transfer adversarial attacks. 
To better tailor the assessment, we proposed ESMA and BEM-ESMA as evaluation measures.
Specifically, we introduced the concept of local sample density and empirically and theoretically confirmed that adversarial perturbations targeting HSDR significantly enhance transferability, resulting in ESMA.
Furthermore, for broader applicability and incorporating information bottleneck theory, we introduced BEM-ESMA.
Both ESMA and BEM-ESMA achieve outstanding performance.
Using ESMA and BEM-ESMA, we conducted security risk assessments, evaluating erasure and tampering risks across code length and model architecture. Our findings are as follows: 
1) Deep watermarking faces risks of erasure and tampering, and can even be manipulated to specific target watermarks. In specific scenarios, invisible deep watermarks are susceptible to non-targeted attacks and can be completely erased.
2) A trade-off exists between the transformation robustness and inherent security of deep watermarks.
3) Architectures with strong transformation robustness are more vulnerable to being targeted for tampering attempts.
4) A trade off exists between the encoding bits and adversarial robustness against erased attacks.

We aim for our research to trigger heightened interest and thorough investigation into the intrinsic security of deep watermarking technologies, especially in advancing more trustworthy deep watermark injection techniques. A variety of intriguing questions await exploration in the future, including learning methods and model architectures that collectively ensure robust input transformation and watermark security.

\bibliographystyle{IEEEtran}
\bibliography{IEEEabrv,ref}

\end{document}